\documentclass[pra,10pt,a4paper,twocolumn,eqsecnum,showpacs]{revtex4}%
\usepackage{amsfonts}
\usepackage{amsmath}
\usepackage{amssymb}
\usepackage{graphicx}%
\begin{document}
\title[ ]{Quantum feedback cooling of a single trapped ion in front of a mirror}
\author{V. Steixner}
\author{P. Rabl}
\author{P. Zoller}
\affiliation{Institute for Theoretical Physics, University of Innsbruck, 6020 Innsbruck, Austria}
\affiliation{Institute for Quantum Optics and Quantum Information of the Austrian Academy
of Sciences, 6020 Innsbruck, Austria}
\date{\today}

\begin{abstract}
We develop a theory of quantum feedback cooling of a single ion
trapped in front of a mirror. By monitoring the motional sidebands
of the light emitted into the mirror mode we infer the position of
the ion, and act back with an appropriate force to cool the ion.
We derive a feedback master equation along the lines of the
quantum feedback theory developed by Wiseman and Milburn, which
provides us with cooling times and final temperatures as a
function of feedback gain and various system parameters.
\end{abstract}
\maketitle

\section{Introduction}

Laser cooling and trapping of single ions
\cite{BlattRMP,WinelandCooling,EITCooling} is one of the highlights in the
development of quantum optics during the last two decades. Single trapped ions
are a laboratory paradigm of a quantum system, which can be prepared and
controlled on the single quantum level, and whose time evolution can be
monitored continuously by observing the scattered light in photodetection or
homodyne measurements \cite{BlattRMP}. By continuous observation of a single
quantum system \cite{CarmichaelBook,GZ} we learn the state of the system, as
described by a conditional system density matrix $\rho_{c}(t)$, and based on
this knowledge we can act back on the system, giving rise to quantum feedback
control of the system of interest
\cite{WM_coll,Feedback_Squeezing,FB_CQED,FB_Ion,FB_Meso}. In the present paper
we present a theory of \emph{quantum feedback cooling of a single trapped
ion}: by extracting from the scattered light the position of the ion in the
trap, we implement a feedback loop on the system in the form of a damping
force with the purpose of cooling the ion motion in the trap. Development of
this theory is not only of fundamental interest in quantum optics, but the
particular setup studied is motivated by ongoing experimental efforts
\cite{BushevThesis,FeedbackCoolingExp} to implement quantum feedback cooling
of single trapped ions in laboratory. Indeed the present theoretical results
provides a quantitative basis for the understanding of these experiments
\cite{FeedbackCoolingExp}.

The particular setup studied in the present paper is a single laser cooled
trapped ion in front of a mirror\cite{MirrorColl}, as illustrated in
Fig.~\ref{fig:physicalsetup}, and motivated by present experiments
\cite{BushevThesis,FeedbackCoolingExp}. A single ion is stored a distance $L$
from a mirror in a harmonic trapping potential. The ion is assumed to be a
two-level system weakly driven near resonance by laser light. Light is
scattered into both the mirror mode, as well as the other other ``background''
modes of the radiation field. By detuning the laser on the red side of the
atomic transition, the ion is laser cooled to a temperature corresponding to
Doppler limit, where the mean occupation of the trap levels is much larger
than one (i.e. far away from the sideband cooling limit to the ground state of
the trap). Motion of the ion adds sidebands of the light scattered into the
mirror mode displaced by the trap frequency. Observing the scattered light of
these motional sidebands allows us to infer the position of the ion in the
sense of continuous measurement theory, and feed back a damping force
proportional to the momentum to implement quantum feedback cooling. In this
paper we will first formulate a continuous measurement theory to read the
position of the trapped ion from the scattered light using the language of
stochastic Schr{\"o}dinger Equations \cite{GZ,CarmichaelBook}. Building on
general quantum feedback theory formulated by Wiseman and Milburn
\cite{WM_coll,Feedback_Squeezing}, we will then derive a quantum feedback
master equation for the motion of the trapped ion. This will allow us to study
the dynamics and limits of quantum feedback cooling.

For the setup studied in this paper the continuous readout of the ion position
is based on light scattering into the mirror mode, with additional photons
scattered into all other ``background modes'' of the radiation field.
Spontaneous emission is intrinsically associated with a momentum recoil of the
ion, which perturbs the ion motion, i.e.~contributes a heating mechanism for
the ion. In a parallel paper \cite{Rabl} we study a quantum feedback scheme
based on a \emph{dispersive readout of the velocity} of the trapped ion to
avoid this unwanted heating. It is based on the large variation of the index
of refraction with the Doppler effect near a \emph{dark state resonance in an
atomic $\Lambda$-system} (based on electromagnetically induced transparency).

The paper is structured as follows. Sec. II presents the basic dynamic
equations for the motion of an ion in front of a mirror. Quantum feedback
equations are formulated in Sec. III, while results are presented in Sec. IV.

\section{Model and Basic Equations}

In this section we will develop the basic equations for continuous measurement
of the photons in the mirror mode of the electromagnetic field. We will start
with a detailed description of our model in terms of a Schr{\"o}dinger
equation for the coupled atom-bath system and the exciting laser. Continuous
measurement theory provides us with a quantum stochastic Schr{\"o}dinger
equation and hence a quantum stochastic master equation in the Lamb-Dicke
limit, where we adiabatically eliminate the excited state from the two-level
atom. We will then derive the photocurrent obtained by detecting mirror mode
photons and the corresponding stochastic master equation for the conditional
density operator in the white noise (diffusive) limit.

\subsection{Single trapped ion in front of a mirror}

\label{sec:basics}

\begin{figure}[ptb]
\centering
\includegraphics[width=.4\textwidth]{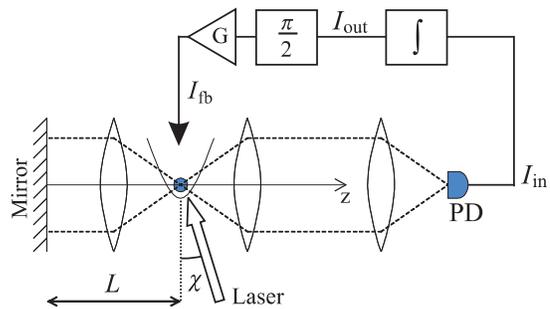}\caption{Physical
Setup: $L$ is the distance between the ion trap center and the mirror, $\chi$
is the incident laser angle. The light is collected in the photodetector PD.
The feedback circuit consists of a bandpass filter, a phase shifter and an
amplifier. The current $I_{\mathrm{fb}}$ is fed back to electrodes creating an
additional potential for the ion. The mirror axis is equal to the z axis.}%
\label{fig:physicalsetup}%
\end{figure}

We consider a single trapped ion which is placed at a distance $L$ from a mirror as indicated in Fig.~\ref{fig:physicalsetup}
\cite{BlattRMP,CiracLaserCooling,Dorner02,EschnerNature}. For the harmonic motion we assume a 1D model in the $z$-direction (identical to the mirror axis). The
harmonic trap has an oscillation frequency $\nu_{T}$, and we denote the destruction (creation) trap operator by $a$ ($a^{\dagger}$). The electronic degrees of
the ion form a two-level atom with atomic transition frequency $\omega_{eg}$, with ground state $\left\vert g\right\rangle $ and excited state $|e\rangle$. We
drive the two-level system with a laser with frequency $\omega_{L}$ which couples the ground to the excited state with the Rabi frequency $\Omega$ and a detuning
from the atomic resonance $\Delta _{L}=\omega_{L}-\omega_{eg}$. The atomic system Hamiltonian can thus be written as
\begin{equation}
H_{\mathrm{sys}}=\nu_{T}a^{\dag}a-\Delta_{L}\left\vert e\right\rangle
\left\langle e\right\vert -\frac{1}{2}\Omega\left(  \mathrm{e}^{\mathrm{i}%
{k}_{\mathrm{eff}}\hat{z}}\left\vert e\right\rangle \left\langle g\right\vert
+\mathrm{h.c.}\right) \label{Hsys}%
\end{equation}
Note that in this paper we set $\hbar=1$. In the interaction term we allow for a laser field incident at an angle $\chi$ with respect to an axis normal to the
$z$-axis. The momentum recoil due to absorption of a laser photon is represented by ${k}_{\mathrm{eff}}\hat{z}=\eta\sin\chi\left(  a+a^{\dag }\right)
\equiv\tilde{\eta}\left(  a+a^{\dag}\right)  $ where the Lamb-Dicke parameter $\eta=2\pi a_{0}/\lambda$ is the ratio of the size of the ground state and the
laser wavelength. Due to the geometry of the system in consideration, the (quantized) electric field consists of two contributions,
$\vec{E}^{\left(  +\right)  }=\vec{E}_{m}^{\left(  +\right)  }+\vec{E}%
_{b}^{\left(  +\right)  }$, where the $\vec{E} _{m}^{\left(  +\right)  }$ denotes the modes restricted by the boundary condition of the mirror and $\vec{E}
_{b}^{\left(  +\right)  }$ the remaining background  modes \cite{Dorner02,EschnerNature}. We adopt a 1D model for the mirror mode and write for the electric
field operator
\begin{equation}
\vec{E}_{m}^{\left(  +\right)  }\left(  z\right)  =\mathrm{i}\int_{0}^{\infty
}d\omega\,\alpha_{\omega}\vec{e}\sin\left(  k(\omega)z\right)  b_{m}(\omega)
\end{equation}
with $\alpha_{\omega}$ a normalization factor for the mode function. The internal states of the atom couple to the vacuum field by an electric dipole
transition. Denoting by $\vec{d}$ the dipole matrix element, and introducing Pauli operator notation for the two level system, $\sigma_{-}=\left\vert
g\right\rangle \left\langle e\right\vert $, the system-bath coupling Hamiltonian is
\begin{equation}
H_{\mathrm{int}}=-\vec{d}\left(  \vec{E}_{b}^{(+)}(\hat{z})+\vec{E}%
_{m}^{\left(  +\right)  }\left(  \hat{z}\right)  \right)  \sigma
_{-}+\mathrm{h.c.}%
\end{equation}

The total Hamiltonian for the ion coupled to the radiation field is
\begin{equation}
H=H_{\mathrm{sys}}+H_{\mathrm{bath}}+H_{\mathrm{int}}.
\end{equation}
Here $H_{\mathrm{bath}}$ is the free Hamiltonian for the radiation field. We
write this Hamiltonian as the sum of a Hamiltonian for the mirror and the
background modes $H_{\mathrm{bath}}=H_{m}+H_{b}$. In our 1D model the mirror
mode Hamiltonian has the form $H_{m}=\int d\omega\,\omega b_{m}^{\dagger
}(\omega)b_{m}(\omega)$ with $b_{m}(\omega)$ photon destruction operators,
satisfying bosonic commutation relations $[b_{m}(\omega),b_{m}^{\dagger
}(\omega^{\prime})]=\delta(\omega-\omega^{\prime})$. Similar expression can be
given for the background modes.

In analyzing this problem we are interested in the situation where the time
delay $\tau_{M}=2L/c$ of the emitted light bouncing from mirror back to the
atom is much shorter than the system time scales, in particular the
spontaneous emission time from the excited state, $\tau_{M}\ll1/\Gamma$, and
the timescales associated with the laser interactions $\tau\ll1/\Omega$,
$1/|\Delta_{L}|$. This justifies the Markov approximation for the emission
into the mirror modes, where we refer to \cite{Dorner02} for a complete analysis.

In the following we will denote the total spontaneous emission rate of the atom by $\Gamma=\Gamma_{m}+\Gamma_{b}.$ Here $\Gamma_{m}=\varepsilon\Gamma$ with
$\varepsilon$ the fraction of the solid angle covered by the lens is the emission rate into mirror mode, and $\Gamma_{b}=(1-\varepsilon)\Gamma$ the emission
rate into the background modes.

\subsection{Quantum Stochastic Schr{\"o}dinger Equation}

The dynamics of our model is summarized in the Schr{\"o}dinger Equation
\begin{align}
|\dot{\Psi}(t)\rangle= &  [-\mathrm{i}H_{\mathrm{sys}}+\label{eq:QSSE}\\
+ &  \sqrt{\Gamma_{m}}\sigma_{-}\sin\left(  k_{eg}(L+\hat{z})\right)
b_{m}^{\dag}(t)+\mathrm{h.c.+}\nonumber\\
+ &  \sqrt{\Gamma_{b}}\int_{-1}^{+1}du\,\sqrt{N(u)}\sigma_{-}\mathrm{e}%
^{-\mathrm{i}uk_{eg}\hat{z}}b_{u}^{\dag}(t)+\mathrm{h.c.}]\left\vert
\Psi(t)\right\rangle \nonumber
\end{align}
We choose to formulate the problem in the language of a \emph{Quantum
Stochastic Schr{\"o}dinger equation} (QSSE) \cite{GZ}, which allows for a
direct connection with continuous measurement of the scattered light, and
provides a direct link to quantum feedback theory developed in the following subsections.

In Eq.~(\ref{eq:QSSE}) $|\Psi(t)\rangle$ is the Schr{\"o}dinger state vector
of the combined atom-field system, i.e. the laser-driven trapped ion including
the mirror and background modes of the radiation field. The first term on the
RHS is the time evolution due to the system Hamiltonian (\ref{Hsys}).

The second and third line describe the interaction of the two-level atom with
the mirror mode and the background modes, respectively. We assume that these
radiation modes are initially in the vacuum state. In writing
Eq.~(\ref{eq:QSSE}) we have transformed to an interaction picture with respect
to the free Hamiltonian of the radiation field $H_{B}$ \cite{GZ}. As a result,
we have introduced bath operators for the mirror mode $b_{m}(t)=1/\sqrt{2\pi
}\int d\omega b_{m}(\omega)e^{-i\omega t}$. In the Markovian limit these
operators satisfy bosonic commutation relations
\begin{equation}
\left[  b_{m}(t),b_{m}^{\dag}(s)\right]  =\delta(t-s).\label{eq:bm}%
\end{equation}
In a Quantum Langevin formulation \cite{GZ} $b_{m}(t)$ represents a quantum
noise operator. Thus the second line of (\ref{eq:QSSE}) describes the emission
of photons by the atom into the mirror mode, with the center of the ion trap
displaced a distance $L$ from the mirror. We note that the motion of the ion
couples to the light via the recoil, as seen by the appearance of $\hat{z}$ in
the mirror mode function. This coupling imparts information of the ion motion
on the light emitted in the mirror mode. In the following subsections we will
analyze this scattered light to continuously monitor the atomic motion, with
the goal of implementing a feedback loop to cool the ion. The coupling
strength to the mirror mode is proportional to the square root of the
spontaneous emission probability into the mirror mode $\Gamma_{m}%
\equiv\varepsilon\Gamma$ with $\varepsilon$ the fraction of the solid angle
(typically $\varepsilon$ is much smaller than one).

The third line in Eq.~(\ref{eq:QSSE}) represents spontaneous emission of the
ion into the background modes. This is a coupling term familiar from the
theory of laser cooling of two-level atoms \cite{CiracLaserCooling}.
Spontaneous emission into the background mode is again associated with a
recoil of the ion motion. In our 1D model for the motion of the trapped ion,
it is the projection of this momentum on the $z$-axis which is the relevant
momentum transfer. Denoting by $\theta$ the angle between the emitted photon
and the $z$-axis, and $u=\cos\theta$, we associate the transition for the
excited state to the ground state including the momentum transfer with the
operator $e^{iuk_{eg}\hat{z}}\sigma_{-}$, where $k_{eg}\equiv\omega
_{eg}/c\approx k_{L} $. Spontaneous photons can be emitted in all directions
into the background modes consistent with the dipole radiation pattern of the
given electronic transition. We denote this (normalized) angular dependence by
$N(u)$. Thus the integral over $u$ in the last line of Eq.~(\ref{eq:QSSE})
realizes photon emission into all of these possible directions. The operators
$b_{u}(t)$ are again photon destruction (or noise) operators associated with
these emission directions. They satisfy commutation relations
\begin{equation}
\lbrack b_{u}(t),b_{u^{\prime}}^{\dag}(s)]=\delta(u-u^{\prime})\delta
(t-s),\label{eq:bu}%
\end{equation}
and commute with the mirror bath operators $b_{m}(t)$ introduced above. The
coupling strength to the background modes is proportional to $\sqrt{\Gamma
_{b}}\equiv\sqrt{(1-\varepsilon)\Gamma}$. For red laser detuning $\Delta
_{L}<0$ the cycle of laser excitation followed by spontaneous emission into
the background mode leads to laser cooling.

\subsection{Ito form of the Quantum Stochastic Schr{\"o}dinger Equation}

To give a meaning to the white noise limit (compare Eqs. (\ref{eq:bm}%
,\ref{eq:bu})), we must interpret the Schr{\"o}dinger equation (\ref{eq:QSSE})
as a quantum stochastic \emph{Stratonovich} equation \cite{GZ}. As usual, it
is more convenient to work with an Ito form, where Wiener noise increments
\begin{equation}
dB_{m,u}(t)=\int_{t}^{t+dt}ds\,b_{m,u}(s)
\end{equation}
\textquotedblleft point to the future\textquotedblright, i.e. are
statistically independent of $|\Psi(t)\rangle$. These Wiener noise increments
satisfy the Ito table
\begin{subequations}
\begin{align}
dB_{m}(t)dB_{m}^{\dagger}(t)  &  =dt,\\
dB_{u}(t)dB_{u^{\prime}}^{\dagger}(t)  &  =\delta(u-u^{\prime})dt,
\end{align}
which follow from Eqs. (\ref{eq:bm},\ref{eq:bu}), the other entries of the Ito
table being zero. The resulting Ito QSSE is
\end{subequations}
\begin{align}
\mathrm{(I)}\quad d\left\vert \Psi(t)\right\rangle =  &  [-\mathrm{i}%
H_{\mathrm{eff}}dt+\sqrt{\Gamma_{m}}C_{m}(\hat{z})dB_{m}^{\dag}\left(
t\right) \label{eq:ItoQSSE}\\
&  +\sqrt{\Gamma_{b}}\int_{-1}^{+1}du\,\sqrt{N(u)}C_{u}(\hat{z})dB_{u}^{\dag
}(t)]\left\vert \Psi(t)\right\rangle \nonumber
\end{align}
Here, we have introduced the \textquotedblleft jump operators\textquotedblright%
\begin{subequations}
\begin{align}
C_{u}(\hat{z})  &  =\mathrm{e}^{-\mathrm{i}uk_{eg}\hat{z}}\sigma
_{-},\label{eq:jump1}\\
C_{m}(\hat{z})  &  =\sin\left(  k_{eg}\left(  L+\hat{z}\right)  \right)
\sigma_{-},\label{eq:jump2}%
\end{align}
which are associated with the emission of a photon in the background modes and
the mirror modes, respectively. Furthermore, we have defined an
\emph{effective non-hermitian system Hamiltonian}
\end{subequations}
\begin{equation}
H_{\mathrm{eff}}=H_{\mathrm{sys}}-\frac{\mathrm{i}}{2}\left[  \Gamma
_{b}+\Gamma_{m}\sin^{2}(k_{eg}(L+\hat{z}))\right]  \left\vert e\right\rangle
\left\langle e\right\vert .\label{eq:Heff}%
\end{equation}
The non-hermitian part of $H_{\mathrm{eff}}$ arises from the Ito correction in
the conversion process. Physically, it corresponds to the radiation damping of
the excited state due to the total radiation field. We also note that the
photon absorption terms have disappeared in Eq.~(\ref{eq:ItoQSSE}) due to
$dB_{m,u}(t)|\Psi(t)\rangle=0$. This follows from our assumption of an initial
vacuum state.

\subsection{Quantum Stochastic Master Equation}

We are interested in the time evolution of our system where the photons
emitted in the mirror mode are detected by a photon counter, while the
background modes remain unobserved. Therefore, we are only interested in the
dynamics of the reduced density operator $\hat{W}\left(  t\right)
\equiv\operatorname{Tr}_{b}\left\{  \left\vert \Psi\left(  t\right)
\right\rangle \left\langle \Psi\left(  t\right)  \right\vert \right\}  $ where
we trace over the background modes of the radiation field. We emphasize that
$\hat{W}\left(  t\right)  $ still contains all the degrees of freedom of the
mirror modes, in addition to the internal and external atomic dynamics.

Using Ito calculus (see Appendix \ref{sec:meq_derivation}) we obtain the
quantum stochastic master equation (QSME)
\begin{align}
\mathrm{(I)}\,\, d\hat{W}\left(  t\right)  =  &  -\mathrm{i}\left( H_{\mathrm{eff}}\hat{W}\left(  t\right)  -\hat{W}\left(  t\right)
H_{\mathrm{eff}}^{\dag}\right)  dt\label{eq:qsme_general}\\
  +&\Gamma_{m}\mathcal{J}\left[  C_{m}(\hat{z})\right]  dB_{m}^{\dag}\left(
t\right)  \hat{W}\left(  t\right)  dB_{m}\left(  t\right) \nonumber\\
  +&\sqrt{\Gamma_{m}}\left(  C_{m}(\hat{z}) dB_{m}^{\dag} \hat W(t) + \hat W(t)C_m^\dag(\hat z) dB_m  \right) \nonumber\\
 +& \Gamma_{b}\int_{-1}^{+1}du\,N(u) \mathcal{J}\left[  C_{u}(\hat{z})\right] \hat{W}\left(  t\right)  dt\nonumber
\end{align}
with $H_{\mathrm{eff}}$ defined in Eq. (\ref{eq:Heff}). For the
\textquotedblleft recycling terms\textquotedblright\ we use the notation
\begin{equation}
\mathcal{J}\left[  c\right]  \rho\equiv c\rho c^{\dag}.
\end{equation}

Before proceeding we note that for $\varepsilon=0$, i.e. no coupling to the
mirror modes, Eq.~(\ref{eq:qsme_general}) reduces to the standard master
equation for 1D laser cooling of a two-level atom \cite{GZ}. In this case
$\hat{W}$ is only an atomic density operator containing the internal and
motional dynamics. For $\varepsilon\neq0$, we still have a stochastic equation
with the mirror bath degrees of freedom included.

\subsection{Adiabatic elimination of the excited state and Lamb-Dicke limit}

We will simplify the above QSSE (\ref{eq:ItoQSSE}) and QSME
(\ref{eq:qsme_general}) with two assumptions. First, we assume weak laser
excitation to the excited state, $\Omega\ll\max\left(  \Gamma,\left\vert
\Delta\right\vert \right)  $. Second, we assume a small Lamb-Dicke parameter
$\eta\equiv2\pi a_{0}/\lambda\ll1$ (tight trap): this allows us to expand the
exponents $\mathrm{e}^{\mathrm{i}k\hat{z}}\equiv\mathrm{e}^{\mathrm{i}%
\eta(a+a^{\dagger})}=1+\mathrm{i}\eta(a+a^{\dagger})+\mathcal{O}(\eta^2)$. Both of these assumptions are well satisfied in present experiments \cite{BlattRMP}

To eliminate the weakly populated excited level, we go back to Eq.~\eqref{eq:ItoQSSE} and expand the state vector $|\Psi(t)\rangle$ into ground state and
excited state components,
\begin{equation}
\left\vert \Psi\left(  t\right)  \right\rangle \equiv\left\vert \psi
_{g}\left(  t\right)  \right\rangle \otimes\left\vert g\right\rangle
+\left\vert \psi_{e}\left(  t\right)  \right\rangle \otimes\left\vert
e\right\rangle \,.\label{eq:decompose}%
\end{equation}
As shown in Appendix \ref{sec:AE_LDL_LC} we can eliminate $|\psi_{e}%
(t)\rangle$ in perturbation theory in the Ito QSSE~(\ref{eq:ItoQSSE}) to
obtain an effective equation for $\left\vert \psi_{g}\left(  t\right)
\right\rangle $. In a similar way as for Eq.~(\ref{eq:qsme_general}) we obtain
a QSME for the partially reduced density operator
\begin{equation}
\hat{w}\left(  t\right)  \equiv\operatorname{Tr}_{b}\left\{  \left\vert
\psi_{g}\left(  t\right)  \right\rangle \left\langle \psi_{g}\left(  t\right)
\right\vert \right\}  ,\label{eq:w}%
\end{equation}
given by
\begin{align}
\mathrm{(I)}\quad d\hat{w}(t) = & -\mathrm{i}\left[  h_{\mathrm{eff}}\hat
{w}(t)-\hat{w}(t)h_{\mathrm{eff}}^{\dag}\right]  dt\label{eq:qsme_lambdicke}\\
&  +\gamma\mathcal{J}\left[  c_{m}(\hat{z})\right]  dB_{m}^{\dag}(t)\hat
{w}(t)dB_{m}(t)\nonumber\\
&  + \sqrt{\gamma} \left(c_m(\hat z) dB_m^\dag(t) \hat w(t) + \hat w(t) c^\dag_m(\hat z) dB_m(t)  \right)\nonumber\\
&+\mathcal{L}_{b}\hat{w}(t) dt.\nonumber
\end{align}
The first three lines give the dynamics of the ion motion coupled to the mirror mode. The fourth line describes the traced-out action of the background mode on
the ion motion, i.e. laser cooling of the ion.

In Eq. (\ref{eq:qsme_lambdicke}) we have defined an effective Hamiltonian
acting only on the motional states of the ion,
\begin{equation}
h_{\mathrm{eff}}=H_{T}-\frac{\mathrm{i}}{2}\gamma c_{m}^{\dag}(\hat{z}%
)c_{m}(\hat{z}).\label{eq:heff}%
\end{equation}
where we expand the eliminated jump operators to second order in the
Lamb-Dicke limit with the center of the trap at $k_{eg}L=\pi/4$:%
\begin{equation}
c_{m}(\hat{z})\approx\frac{1}{\sqrt{2}}\left(  1+\eta\left(  a+a^{\dag
}\right)  -\frac{1}{2}\eta^{2}\left(  a+a^{\dag}\right)  ^{2}\right)  .
\end{equation}
The parameter
\begin{equation}
\gamma=\varepsilon\Gamma\frac{\Omega^{2}}{4}\frac{1}{\Delta_{L}^{2}%
+\frac{\Gamma}{2}}\label{eq:gammamirror}%
\end{equation}
is the optical pumping rate into the mirror mode. The first three lines of Eq.~(\ref{eq:qsme_lambdicke}) thus describe the motional state coupled via laser
excitation followed by spontaneous emission to the mirror mode.

The Liouvillian\ $\mathcal{L}_{b}$ in the fourth line of Eq. (\ref{eq:qsme_lambdicke}) is the standard laser cooling Liouvillian for weak field excitation and
in the Lamb-Dicke limit \cite{BlattRMP,CiracLaserCooling,WinelandCooling},
\begin{align}
\mathcal{L}_{b}\hat{w}\left(  t\right)   &  =A_{-}\mathcal{D}\left[  a\right]
\hat{w}\left(  t\right)  +A_{+}\mathcal{D}\left[  a^{\dag}\right]  \hat
{w}\left(  t\right) \\
&  \equiv\Gamma_{\text{\textrm{eff}}}(N+1)\mathcal{D}\left[  a\right]  \hat
{w}\left(  t\right)  +\Gamma_{\text{\textrm{eff}}}N\mathcal{D}\left[  a^{\dag
}\right]  \hat{w}\left(  t\right) ,\nonumber
\end{align}
where we have used the notation
\begin{equation}
\mathcal{D}\left[  c\right]  \rho\equiv c\rho c^{\dag}-\frac{1}{2}\left(
c^{\dag}c\rho+\rho c^{\dag}c\right)  .
\end{equation}
The rates
\begin{equation}
A_{\pm}=\eta^{2}\frac{\Omega^{2}}{4}\Gamma_{b}\left(  \frac{\sin^{2}\chi }{\left(  \Delta_{L}\mp\nu_{T}\right)  ^{2}+\frac{\Gamma^{2}}{4}}+\frac
{\alpha}{\Delta_{L}^{2}+\frac{\Gamma^{2}}{4}}\right)  .\label{eq:aplusminus}%
\end{equation}
have the meaning of cooling (heating) terms for red laser detuning $\Delta _{L}<0.$ With $\Gamma_{\mathrm{eff}}=A_{-}-A_{+}>0$ and for $\Delta_{L}<0$ we have
\begin{equation}
N=\frac{A_{+}}{A_{-}-A_{+}},
\end{equation}
which is the final mean trap occupation established by laser cooling (alone).
We have also used the abbreviation $\alpha=\int du\,u^{2}N(u)$ for the dipole
transition parameter and $\chi$ is the incident angle of the laser beam. With
these definitions the mirror mode optical pumping rate (\ref{eq:gammamirror})
can be written as $\gamma=\varepsilon N\Gamma_{\mathrm{eff}}/(1+\alpha
)\eta^{2}$, and from $\Gamma_{\mathrm{eff}}\propto\sin^{2}\chi$ and
$N\propto1/\sin^{2}\chi$ we see that this pumping rate is independent from the
angle of the incoming laser beam.

In the following we will study a scenario \cite{FeedbackCoolingExp} where the
laser cooling establishes a steady state with a mean trap occupation $N\gg1$
(i.e. far from the ground state), as represented by the second line in Eq.
(\ref{eq:qsme_lambdicke}). This is the limit of Doppler cooling, which is
obtained if $\Gamma\gg\nu_{T}$. The minimally obtainable steady state energy
in this limit is $\hbar\Gamma(\alpha+1)/2$. By observing the spontaneous
emission into the mirror mode (see first two lines of Eq.
(\ref{eq:qsme_lambdicke})), we will infer the position of the atom to apply a
feedback loop to cool the system (far) below the laser cooling limit.

\begin{figure}[ptb]
\centering
\includegraphics[width=.23\textwidth]{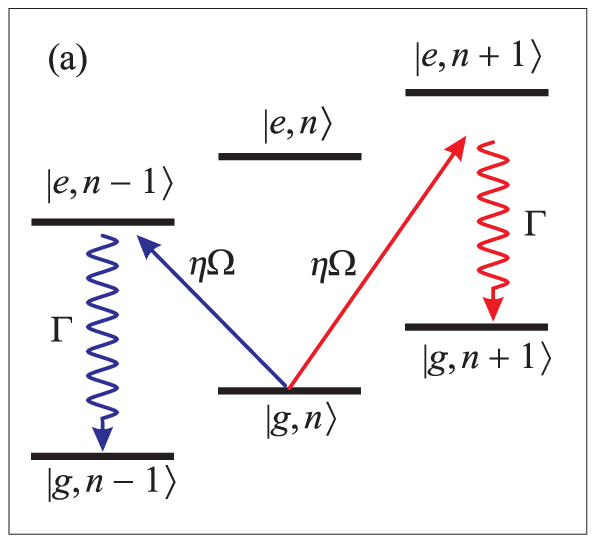}
\includegraphics[width=.23\textwidth]{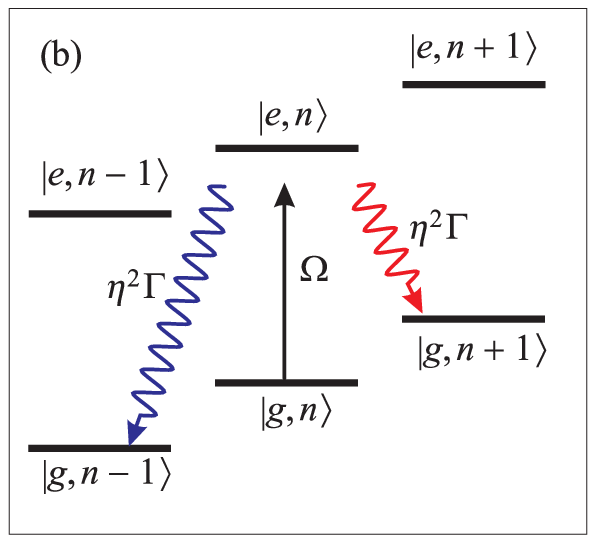}\caption{Contributing
processes in the master equation at low intensity. (a) shows the cooling and
heating terms due to the coupling of the laser to the motion and (b) shows the
diffusion terms due to spontaneous emission \cite{ResFluorescence}.}%
\label{fig:coolheatdiffusion}%
\end{figure}

For completeness we note that in the case where the mirror mode is not
observed, the reduced system density operator $\rho(t)\equiv\operatorname{Tr}%
_{m}\left\{  \hat{w}(t)\right\}  $ obeys the master equation
\begin{align}
\dot{\rho}(t) &  =-\mathrm{i}\left[  H_{T},\rho(t)\right]  +\gamma
\mathcal{D}\left[  c_{m}(\hat{z})\right]  \rho(t)+\mathcal{L}_{b}%
\rho(t)\label{eq:tracedmeq}\\
&  \equiv-\mathrm{i}\left[  H_{T},\rho(t)\right]  +\mathcal{L}_{\mathrm{LC}%
}\rho(t)\equiv\mathcal{L}_{0}\rho(t),\nonumber
\end{align}
which contains the dynamics from the free ion motion, and the dissipative
dynamics from the emission into the mirror mode and laser cooling. In a second
order expansion in terms of $\eta$, we have
\begin{equation}
\mathcal{D}\left[  c_{m}\left(  \hat{z}\right)  \right]  =\eta^{2}\cos
^{2}(k_{eg}L)\,\mathcal{D}\left[  a+a^{\dag}\right]  +\mathcal{O}\left(
\eta^{3}\right)
\end{equation}
which, multiplied by $\gamma$, is typically much smaller than $\Gamma_{\mathrm{eff}}N$ and thus the corrections in the heating and cooling rates will be
neglected here.

\subsection{Continuous observation of the mirror mode\label{sec:cond_meq}}

We measure the photons emitted into the mirror modes by a photon counter as
shown in Fig.~\ref{fig:physicalsetup}. We denote by $N_{c}(t)$ the number of
photon counts at time $t$. A particular count trajectory is characterized the
photon detection times $t_{1},t_{2},\ldots$. Our knowledge of the state of the
system, given by the internal and external degrees of the ion, for a given
count trajectory is represented by a conditional density matrix $\rho
_{c}\left(  t\right)  $ \cite{GZ}.

Given the state of the system at time $t$, $\rho_{c}\left(  t\right)  $, the
detection of a mirror mode photon in a time interval $\left(  t,t+dt\right]  $
is associated with a quantum jump of the atom described by%
\begin{equation}
\rho_{c,\mathrm{jump}}\left(  t+dt\right)  =\frac{\mathcal{J}\left[
c_{m}\right]  \rho_{c}\left(  t\right)  }{\operatorname{Tr}\left\{
\mathcal{J}\left[  c_{m}\right]  \rho_{c}\left(  t\right)  \right\}
}\label{eq:jump}%
\end{equation}
where according to (\ref{eq:jump2}) the atom returns to from the excited state
to the ground state, and momentum is transferred to the ion motion in
accordance with the mirror mode function. In the case of no observed photon,
the system evolves with the effective non trace-preserving Liouvillian $L_{0}
$%
\begin{equation}
\rho_{c,\mathrm{no\ jump}}\left(  t+dt\right)  =\left(  1+L_{0}dt\right)
\rho_{c}\left(  t\right) \label{eq:nojump}%
\end{equation}
where
\[
L_{0}\rho\equiv-\mathrm{i}\left[  h_{\mathrm{eff}}\rho-\rho h_{\mathrm{eff}%
}^{\dag}\right]  +\mathcal{L}_{b}\rho
\]
and $h_{\mathrm{eff}}$ is defined in Eq.~(\ref{eq:heff}). The expected number
of counts in the interval $(t,t+dt]$ is with $dN_{c}(t)=N_{c}\left(
t+dt\right)  -N_{c}(t)$
\begin{equation}
\left\langle dN_{c}(t)\right\rangle =p_{\mathrm{emission}}^{\left(
t,t+dt\right]  }=\gamma\operatorname{Tr}_{\mathrm{sys}}\left\{  \mathcal{J}%
\left[  c_{m}\right]  \rho_{c}\left(  t\right)  \right\}
dt\label{eq:meandNgeneral}%
\end{equation}
In view of
$dN_{c}(t)=0$ or $1$,  for this point process we have the Ito table $dN_{c}%
^{2}\left(  t\right)  =dN_{c}\left(  t\right)  ~$ and $dN_{c}\left(  t\right) dt=0$.

We can summarize the above \emph{a posteriori} time evolution in an Ito
stochastic Schr{\"o}dinger equation (see. eg. \cite{GZ})
\begin{align}
\mathrm{(I)}\quad\rho_{c}\left(  t\right)  = &  \mathcal{L}_{0}\rho_{c}\left(
t\right)  dt+\label{eq:condmastereq}\\
&  +\left(  \frac{\mathcal{J}\left[  c_{m}\right]  \rho_{c}\left(  t\right)
}{\operatorname{Tr}_{\mathrm{sys}}\left\{  \mathcal{J}\left[  c_{m}\right]
\rho_{c}\left(  t\right)  \right\}  }-\rho_{c}\left(  t\right)  \right)
\times\nonumber\\
&  \times\left(  dN_{c}\left(  t\right)  -\gamma\operatorname{Tr}%
_{\mathrm{sys}}\left\{  \mathcal{J}\left[  c_{m}\right]  \rho_{c}\left(
t\right)  \right\}  dt\right) \nonumber
\end{align}
where $\mathcal{L}_{0}$ is defined in (\ref{eq:tracedmeq}). This equation gives the time evolution of the conditional density matrix of the ion $\rho_{c}\left(
t\right)  $ for a particular count trajectory. Not observing, i.e. tracing over the mirror mode, is equivalent to taking the ensemble average over all count
trajectories in (\ref{eq:condmastereq}). In this case, we recover the master equation $\dot{\rho}\left(  t\right)  =\mathcal{L}_{0}\rho\left( t\right)  $ for
the \emph{a priori }dynamics \cite{GZ}.

\subsection{Diffusion approximation}

In the previous subsection we considered photon counting of the light emitted
in the mirror modes, and the associated time evolution of the system described
by the condition density operator $\rho_{c}\left(  t\right)  $. We are
interested in learning the motion (position) of the atom from the scattered
light in the sense of continuous measurement. The goal is to use this
information to control the motion of the atom, and eventually act back on the
atom to cool it.

The scattered light of a weakly driven trapped atom \cite{ResFluorescence}
consists of (i) a strong elastic component at the frequency of the driving
laser (see vertical transitions in Fig. \ref{fig:coolheatdiffusion}), and (ii)
weak motional sidebands at the trap frequency $\nu_{T}$ suppressed by the
Lamb-Dicke parameter $\eta$. The information on the motion of the atom is
encoded in the \textquotedblleft motional sidebands\textquotedblright. We find
it convenient to formulate the problem in a way, where we focus directly on
the contributions of these sidebands to the photon count signal. The physical
picture is that the elastic component acts like a \textquotedblleft(strong)
local oscillator\textquotedblright\ which beats with the \textquotedblleft%
(weak) light emitted from the sidebands\textquotedblright. This situation is
reminiscent of homodyne measurements in quantum optics
\cite{GZ,CarmichaelBook}, and will lead in the following to a description in
terms of a \emph{diffusive stochastic process} rather than a point process
associated with the photon counting described above. The formal expansion
parameter is $\eta\ll1$ (Lamb-Dicke limit).

From the previous subsection we know that the mean number of photon counts in
$\left(  t,t+dt\right]  $ is
\begin{equation}
\left\langle dN_{c}(t)\right\rangle \overset{kL=\pi/4}{=}\frac{1}{2}\gamma
dt+\gamma\eta\left\langle a+a^{\dag}\right\rangle _{c}(t)dt+O(\eta
^{2}).\label{eq:meannumberdN}%
\end{equation}
The first term is elastic scattering. The second term, which is first order in
$\eta$, is proportional to $\tilde{z}\equiv a+a^{\dag}$, i.e. includes
information on the ion motion. Here and in the following we take the center of
the trap to be on the slope of the standing wave, i.e. $k_{eg}L=\pi/4$.

Following the analysis of homodyne detection \cite{GZ,CarmichaelBook}, we
split the stochastic variable $dN_{c}(t)$ into a deterministic and a remaining
stochastic part, thus defining $dY_{c}\left(  t\right)  $,
\begin{equation}
dN_{c}(t)\equiv\frac{1}{2}\gamma dt+\eta dY_{c}\left(  t\right)
\label{eq:dYdefinition}%
\end{equation}
and we can show (cf. Appendix \ref{sec:homodyne_diffusion}) that $dY_{c}(t)$
is a Gaussian stochastic variable with non-zero mean, i.e.
\[
dY_{c}(t)=\sqrt{\gamma/2}/\eta~dW(t)+\gamma\left\langle \tilde{z}\right\rangle
_{c}(t)dt
\]
with $dW(t)$ a Wiener increment satisfying $dW^2(t)=dt$.

This leads us to define a photocurrent where we subtract the large constant
contribution from the elastic scattering process,%

\begin{align}
I_{c}(t) &  =\eta\frac{dY_{c}(t)}{dt}\label{eq:homodyne_current_in}\\
&  =\gamma\eta\left\langle \tilde{z}\right\rangle _{c}(t)+\sqrt{\frac{\gamma
}{2}}\xi\left(  t\right)  .\nonumber
\end{align}
with $\xi\left(  t\right)  $ Gaussian white noise \ $\left\langle \xi\left(
t\right)  \xi\left(  t^{\prime}\right)  \right\rangle =\delta\left(
t-t^{\prime}\right)  $ (shot noise). We see that $I_{c}(t)$ follows
$\langle\tilde{z}\rangle_{c}(t)$ and thus represents a continuous measurement
of the position of the ion. The information on the motion is contained in the
sidebands of the current, i.e. in the frequency components centered around
$\pm\nu_{T}$.

In the diffusive approximation the conditional density matrix $\rho_{c}(t)$
\cite{GZ,CarmichaelBook} obeys
\begin{equation}
\mathrm{(I)}\quad d\rho_{c}(t)=\left[  \mathcal{L}_{0}dt+\sqrt{\frac{\gamma
}{2}}dW(t)~\mathcal{H}_{m}\right]  \rho_{c}(t)\label{eq:rhocond_homodyne_ito}%
\end{equation}
where%
\begin{equation}
\mathcal{H}_{m}\rho_{c}(t)=2\eta\left(  \tilde{z}\rho_{c}(t)+\rho_{c}%
(t)\tilde{z}-2\left\langle \tilde{z}\right\rangle _{c}(t)\ \rho_{c}(t)\right)
\end{equation}
and Eq. (\ref{eq:rhocond_homodyne_ito}) is derived from (\ref{eq:condmastereq}%
)\ \ in the diffusive limit $\eta\ll1$ (cf. Appendix
\ref{sec:homodyne_diffusion}).

\section{Quantum Feedback Cooling}

In the previous section we have reformulated the continuous observation of the
ion motion through spontaneous light scattering into mirror modes in a form
reminiscent of homodyne detection. This will allow us below to study feedback
cooling of trapped ions building on the \emph{Wiseman-Milburn theory of
quantum feedback} \cite{WM_coll,Feedback_Squeezing}.

In Eq. (\ref{eq:homodyne_current_in}) we have obtained a current which is
proportional to the mean value of the \emph{position} of the atom. We want to
use this information to feed back an appropriate force proportional to the
\emph{momentum} to damp the motional state of the atom
\cite{BushevThesis,FeedbackCoolingExp}. The information about the position is
encoded in the motional sidebands of the current. In a harmonic trap of known
frequency any combination of the average position and momentum can be obtained
by shifting the sideband current by a phase of $\phi$, if the trap frequency
is much faster than any other (cooling) timescale in the problem (weak
coupling limit). This phase $\phi$ can be controlled electronically, and for
$\phi=\pi/2$, the shifted current follows the momentum. A force, which is
proportional to this current, can damp the motion of the ion.

\subsection{Feedback current}

\begin{figure}[ptb]
\centering
\includegraphics[width=.37\textwidth]{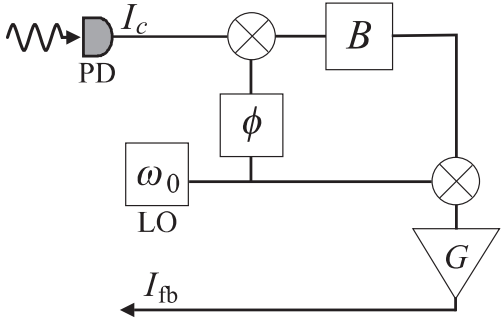}\caption{Electronic
Setup of the Feedback loop as modeled in this paper. PD is the photodetector,
LO is the local oscillator, the $B$-box is the band pass (BP) filter and the
$\phi$-box is the phase shift. The LO signal is mixed to the current and
subtracted after the BP filter.}%
\label{fig:electronics}%
\end{figure}We model the feedback circuit as shown in
Fig.~\ref{fig:electronics}. First, the signal $I_{c}(t)$ given by
Eq.~(\ref{eq:homodyne_current_in}) is mixed with a local oscillator of
frequency $\omega_{0}\approx\nu_{T}$ to shift the signal of the motional
sideband to zero frequency. Then the current is sent through a band pass
filter of width $B$ to cut off rapidly oscillating terms. The filter is
described by a filter function $Z(\omega)$, centered around zero frequency. At
the end the signal is mixed again with the local oscillator and amplified by a
factor $G$. The feedback current can then be written as%
\begin{equation}
I_{\mathrm{fb},c}\left(  t\right)  =G\cos(\omega_{0}t)\int_{-\infty}^{t}%
d\tau\tilde{Z}(t-\tau)\cos\left(  \omega_{0}\tau+\phi\right)  I_{c}(\tau),
\end{equation}
where $\tilde{Z}\left(  \tau\right)  $ is the Fourier transform of the band
pass function $Z(\omega)$. The feedback Hamiltonian is specified in the next subsection.

To evaluate the expression for the current, it is convenient to change to a
basis which is rotating with the frequency of the local oscillator $\omega
_{0}$ by applying the unitary transformation $U\equiv\exp(-\mathrm{i}%
\omega_{0}a^{\dag}at)$. The evolution timescale of the density operator in
this new frame, $\tilde{\rho}_{c}(t)\equiv U\rho_{c}(t)U^{\dag}$ is determined
by the detuning $\delta=\omega_{0}-\nu_{T}$ and the cooling rates
$G\gamma, \Gamma_{\mathrm{eff}}$. Under the assumption, that these frequencies
are smaller than the filter bandwidth $B$, the feedback current is given by%
\begin{equation}
I_{\mathrm{fb},c}\left(  t\right)  =G\left[  \gamma\eta\left\langle X_{\phi
}\right\rangle _{c}^{I}(t)+\sqrt{\frac{\gamma}{2}}\Xi\left(  t\right)
\right]  \cos\left(  \omega_{0}t\right)  .\label{eq:feedback_current}%
\end{equation}
The first term in this expression, $\left\langle X_{\phi}\right\rangle
_{c}^{I}\equiv\operatorname{Tr}_{\mathrm{sys}}\{X_{\phi}\tilde{\rho}_{c}(t)\}$
is the slowly varying expectation value of the quadrature component
\begin{equation}
X_{\phi}\equiv a\mathrm{e}^{\mathrm{i}\phi}+a^{\dag}\mathrm{e}^{-\mathrm{i}%
\phi}\label{eq:xphi}%
\end{equation}
(in the rotating frame). The second contribution in
Eq.~(\ref{eq:feedback_current}) is defined as
\begin{equation}
\Xi(t)\equiv\int_{-\infty}^{t}d\tau\,\cos\left(  \omega_{0}\tau+\phi\right)
\tilde{Z}(t-\tau)\xi\left(  \tau\right)  .\label{eq:filtered_noise_definition}%
\end{equation}
It describes the noise which passes through the feedback circuit. The
stochastic mean of $\Xi(t)$ is zero due to the vanishing mean of the white
noise variable $\xi(t)$, and the correlation function is given by%
\begin{equation}
\left\langle \Xi(t)\Xi(t^{\prime})\right\rangle \approx\delta_{B}\left( t-t^{\prime}\right)  +\mathcal{O}\left(  \frac{B}{\omega_{0}}\right).
\end{equation}
Here $\delta_{B}(t-t^{\prime})$ denotes a delta-function for functions which vary on a slow timescale much larger than $B^{-1}$.

Thus for a clear separation of timescales,
\begin{equation}
G\gamma,\delta,\Gamma_{\mathrm{eff}}\ll B\ll\omega_{0},\nu_{T},
\end{equation}
the current given in Eq.~(\ref{eq:feedback_current}) is proportional to the
slowly varying expectation value of $X_{\phi}$, and has a noise term which is
delta-correlated on a timescale of the system evolution in the rotating frame.

\subsection{Quantum Feedback Dynamics}

The feedback current (\eqref{eq:feedback_current}) for $\phi=-\pi/2$ is proportional to the slowly varying momentum of the particle. For the cooling of the ion
motion, we apply a linear force which is proportional to the the feedback current (\ref{eq:feedback_current}). For a trapped ion, this can be realized by
applying a voltage on the trap electrodes, which leads to a displacement of the trap center. The effect of the feedback force is given by
the interaction picture Hamiltonian%
\begin{equation}
H_{\mathrm{fb}}=I_{\mathrm{fb},c}(t-\tau)\tilde{z}_{I}(t).
\end{equation}
In this equation, $\tilde{z}_{I}(t)\equiv U^{\dag}\tilde{z}U$ is proportional to the position operator in the interaction picture, while $\tau$ denotes the
finite time delay in the feedback loop, which we require to be much smaller than the trap frequency $\tau\ll1/\nu_{T}$. The master equation
(\ref{eq:rhocond_homodyne_ito}) has to be complemented with the feedback term,%
\begin{equation}
\mathrm{(S)}\quad\left[  d\tilde{\rho}_{c}(t)\right]  _{\mathrm{fb}%
}=I_{\mathrm{fb},c}(t-\tau)\left(  -\mathrm{i}\right)  \left[  \tilde{z}%
_{I}(t),\tilde{\rho}_{c}(t)\right]
dt.\label{eq:rhocond_feedback_stratonovich}%
\end{equation}
which has to be interpreted as a \emph{Stratonovich} stochastic differential equation \cite{Feedback_Squeezing}. For the slow dynamics of the density matrix in
the rotating frame, we can make a rotating wave approximation and neglect rapidly rotating terms $\sim\exp (\pm2\mathrm{i}\omega_{0}t)$. The filtered noise
(\ref{eq:filtered_noise_definition}) is delta-correlated on timescales slower
than $B^{-1}$, thus we have the coarse grained evolution of the density matrix%
\begin{align}
\mathrm{(S)}\quad\left[  d\tilde{\rho}_{c}(t)\right]  _{\mathrm{fb}}  &
=\frac{G}{2}\gamma\eta\left\langle X_{\phi}\right\rangle _{c}^{I}%
(t-\tau)dt\mathcal{K}\tilde{\rho}_{c}(t)\label{eq:rhocond_fb_s_rwa}\\
& +\frac{G}{2}\sqrt{\frac{\gamma}{2}}dW_{\Xi}(t-\tau)\mathcal{K}\tilde{\rho
}_{c}(t),\nonumber
\end{align}
with the feedback operator
\begin{equation}
\mathcal{K}\tilde{\rho}_{c}(t)\equiv-\mathrm{i}\left[  \tilde{z},\tilde{\rho
}_{c}(t)\right] \label{eq:superk}%
\end{equation}
and the \textquotedblleft slow\textquotedblright\ Wiener increment $dW_{\Xi
}\left(  t\right)  \equiv\Xi\left(  t\right)  dt$.

The total evolution of the system is determined by the conditioned master
equation (\ref{eq:rhocond_homodyne_ito}) plus the contribution from the
feedback loop (\ref{eq:rhocond_fb_s_rwa}). To combine the two equations, we
have to convert Eq.~(\ref{eq:rhocond_fb_s_rwa}) from Stratonovich to Ito form.
The total conditioned evolution is%
\begin{align}
\mathrm{(I)}\quad d\tilde{\rho}_{c}(t) &  =\mathcal{\tilde{L}}_{0}\tilde{\rho
}_{c}+\sqrt{\frac{\gamma}{2}}\mathcal{H}dW(t)\tilde{\rho}_{c}(t)\nonumber\\
&  +\left(  \frac{G}{2}\gamma\eta\left\langle X_{\phi}\right\rangle _{c}%
^{I}(t-\tau)dt+\frac{G^{2}}{16}\gamma\mathcal{K}dt+\right. \nonumber\\
&  \left.  +\frac{G}{2}\sqrt{\frac{\gamma}{2}}dW_{\Xi}(t-\tau)\right)
\mathcal{K}\tilde{\rho}_{c}(t),\label{eq:rhocond_feedback_ito}%
\end{align}
where
\begin{equation}
\mathcal{\tilde{L}}\tilde{\rho}_{c}\equiv\mathcal{L}_{\mathrm{LC}}\tilde{\rho
}_{c}-\mathrm{i}[\delta a^{\dag}a,\tilde{\rho}_{c}]\label{eq:ltilde}%
\end{equation}
(cf. Eq.~(\ref{eq:tracedmeq})) is the laser cooling Liouvillian in the
rotating frame.

Because the exact photocurrent can not be kept track of in experiments,
Eq.~(\ref{eq:rhocond_feedback_ito})\ is of limited use. The goal is to derive
an equation for the ensemble averaged density operator. We follow the
derivation given by Wiseman and Milburn in \cite{Feedback_Squeezing}, where
the measured current is fed back directly, and adopt it for our model.
Assuming that the state at time $t-\tau$ and all previous times is known, we
take the ensemble average $E[\cdot]$ of Eq.~(\ref{eq:rhocond_feedback_ito})
over the trajectories in $(t-\tau,t]$. We then formally divide by $dt$ and for
convenience redefine $\rho(t)\equiv E[\tilde{\rho}_{c}(t)]$:%
\begin{align}
\mathrm{(I)}\quad\dot{\rho}(t)= &  \mathcal{\tilde{L}}\rho(t)+\frac{G}%
{2}\gamma\eta\left\langle X_{\phi}\right\rangle _{c}^{I}(t-\tau)\mathcal{K}%
\rho(t)\label{eq:ensembleaverage_incomplete}\\
+ &  \frac{G}{2}\sqrt{\frac{\gamma}{2}}\mathcal{K}E\left[  \Xi(t-\tau
)\tilde{\rho}_{c}(t)\right]  +\frac{G^{2}}{16}\gamma\mathcal{K}^{2}%
\rho(t).\nonumber
\end{align}
The density matrix $\rho(t)$ is still conditioned on the evolution up to time
$t-\tau$, but not conditioned on trajectories in $(t-\tau,t]$. The ensemble
average $E[\langle X_{\phi}^{I}\rangle_{c}(t-\tau)\tilde{\rho}_{c}(t)]$
factorizes because $\rho_{c}(t-\tau)$ is assumed known. Under the Markov
approximation, we let $\tau$ go to zero, while due to the coarse graining of
the time evolution in Eq.\thinspace(\ref{eq:rhocond_feedback_stratonovich}),
$dt$ will still be larger than this small delay. An expansion in $\tau$ yields%
\begin{align}
\tilde{\rho}_{c}(t) &  =\left[  1+\mathcal{O}\left(  \tau\right)  \right]
\tilde{\rho}_{c}(t-\tau+dt)=\\
&  =\left[  1+\mathcal{O}\left(  \tau\right)  \right]  \left[  1+\sqrt
{\frac{\gamma}{2}}dW(t-\tau)\mathcal{H}\right]  \tilde{\rho}_{c}%
(t-\tau).\nonumber
\end{align}
We now can evaluate the remaining ensemble average in
Eq.~(\ref{eq:ensembleaverage_incomplete}) because $dW(t-\tau)$ is
stochastically independent from $\tilde{\rho}_{c}(t-\tau)$. We obtain
\begin{align}
E\left[  \Xi(t-\tau)\tilde{\rho}_{c}(t)\right]  = &  \sqrt{\gamma}%
\mathcal{H}E\left[  \Xi(t-\tau)\xi(t-\tau)\right]  \rho
(t)\label{eq:ensembleaveragexi}\\
\simeq &  \sqrt{\frac{\gamma}{2}}\eta\big(X_{\phi}\rho(t)+\rho(t)X_{\phi
}\nonumber\\
&  -2\left\langle X_{\phi}\right\rangle _{c}^{I}(t-\tau)\rho
(t)\big) ,\nonumber
\end{align}
and thus the term in the last line, a conditional expectation value, cancels
with the second term on the right hand side of
Eq.~(\ref{eq:ensembleaverage_incomplete}). In going from the first to the
second line in Eq.~(\ref{eq:ensembleaveragexi}) we have dropped terms
$\sim\exp(\pm\mathrm{i}\omega_{0}t)$.

With this last step, we can finally evaluate
Eq.~(\ref{eq:ensembleaverage_incomplete}) and write down the \emph{quantum
feedback master equation} (compare for the motional degrees of freedom:%
\begin{equation}
\dot{\rho}=\mathcal{\tilde{L}}\rho+\frac{G}{4}\gamma\eta\mathcal{K}\left(
X_{\phi}\rho+\rho X_{\phi}\right)  +\frac{G^{2}}{16}\gamma\mathcal{K}^{2}%
\rho.\label{eq:masterequation_final}%
\end{equation}
The first term on the right hand side $\mathcal{\tilde{L}}$ is the laser
cooling Liouvillian (\ref{eq:ltilde}) in the rotating frame. The second term
with $\mathcal{K}$ given in Eq.~(\ref{eq:superk}) in the master equation is
the feedback term. It acts back on the system and is responsible for cooling
if we choose the parameters $\delta$ and $\phi$ appropriately. The last term
in the master equation is a diffusive term of the form of a double commutator.

\section{Results\label{sec:results}}

In the last section we have shown that for a separation of timescales
$\delta,\Gamma_{\mathrm{eff}}\ll B\ll\omega_{0},\nu_{T}$ we obtain an
unconditioned (non-selective) master equation for the motional density matrix
in the rotating frame. By inserting the definitions of $\mathcal{\tilde L}$
and $\mathcal{K}$ the master equation reads%
\begin{align}
\dot{\rho}  & =-\mathrm{i}\delta\lbrack a^{\dag}a,\rho]+A_{-}\mathcal{D}%
[a]+A_{+}\mathcal{D}[a^{\dag}]+\label{meq_expanded}\\
& -\mathrm{i}\frac{G}{4}\gamma\eta\lbrack\tilde{z},X_{\phi}\rho+\rho X_{\phi
}]-\frac{G^{2}}{16}\gamma\lbrack\tilde{z},[\tilde{z},\rho]].\nonumber
\end{align}
We have used the previously introduced variables $\tilde z=a+a^{\dag}$ and
$X_{\phi}=a\mathrm{e}^{\mathrm{i}\phi}+a^{\dag}\mathrm{e}^{-\mathrm{i}\phi}%
$. In the first line of Eq.~(\ref{meq_expanded}) we recover the master equation for laser cooling, with the corresponding heating and cooling rates $A_{\pm}$
given in Eq.~(\ref{eq:aplusminus}). The second line describes the effect of
the feedback loop, where $\gamma=\varepsilon N \Gamma_{\mathrm{eff}}%
/(1+\alpha)\eta^{2}$ is the emission rate in the mirror mode and $G$ is the gain parameter amplifying the feedback current. The first term in the second line
depends on the phase shift $\phi$ and as we will show below, leads to the expected damping for $\phi=-\pi/2$. The second term arises from the noise in the
feedback current and leads to a momentum diffusion, i.e. heating.

We will derive solutions of the feedback master equation
(\ref{eq:masterequation_final}), which is bilinear in the position and
momentum $\hat{z}$ and $\hat{p}_{z}$. It is convenient to use a Wigner
function representation \cite{GZ} of the density matrix. This gives rise to a
Fokker-Planck equation \cite{Risken} for the Wigner function $W(\bar{z}%
,\bar{p},t)$ with dimensionless position and momentum variables $x_{1}%
\equiv\bar{z}=z\sqrt{m\nu_{T}/2}$ and $x_{2}\equiv\bar{p}=p_{z}/\sqrt
{2m\nu_{T}}$,
\begin{align}
\frac{\partial W(\bar{z},\bar{p},t)}{\partial t}= &  \sum_{i,j}\kappa
_{ij}\frac{\partial}{\partial x_{i}}\left(  x_{j}W(\bar{z},\bar{p},t)\right)
+\label{eq:FPE}\\
&  +\sum_{i,j}D_{ij}\frac{\partial^{2}W(\bar{z},\bar{p},t)}{\partial
x_{i}\partial x_{j}}.\nonumber
\end{align}
The $\kappa_{ij}$ are independent of the phase space variables and $D_{ij}$ is
diagonal, thus Eq.\thinspace(\ref{eq:FPE}) describes an Ornstein-Uhlenbeck
process \cite{Risken} with drift matrix
\begin{equation}
\kappa=\frac{\Gamma_{\mathrm{eff}}}{2}\left(
\begin{array}
[c]{cc}%
1 & -2\tilde{\delta}\\
2G\eta\tilde{\gamma}\cos\phi+2\tilde{\delta} & 1-2G\eta\tilde{\gamma}\sin\phi
\end{array}
\right) \label{eq:kappamatrix}%
\end{equation}
and the diagonal terms of the diffusion matrix%
\begin{equation}
(D_{11},D_{22})=\frac{\Gamma_{\mathrm{eff}}}{8}\left(  2N+1,2N+1+\frac{1}%
{2}G^{2}\tilde{\gamma}\right)  .\label{eq:diffusionterm}%
\end{equation}
Here we have introduced the dimensionless detuning $\tilde{\delta}\equiv
\delta/\Gamma_{\mathrm{eff}}$ and decay rate $\tilde{\gamma}\equiv
\gamma/\Gamma_{\mathrm{eff}}$ normalized with respect to the width of the
sidebands. The Gaussian Wigner function is uniquely determined by it's first
and second position and momentum moments, and we will use the notation
\begin{equation}
\left\langle \bar{z}^{r}\bar{p}^{s}\right\rangle _{W}\equiv\int d\bar{z}%
d\bar{p}\,\bar{z}^{r}\bar{p}^{s}W(\bar{z},\bar{p},t),
\end{equation}
which equals the symmetric expectation value of the corresponding operators.
The bilinearity of Eq.~(\ref{eq:masterequation_final}) with respect to
position and momentum gives rise to a closed set of equations for the first
and second moments of the Wigner function individually and are given in
Appendix \ref{sec:fpe_solution}.

We are interested in the motional energy of the ion, which is related to the
expectation value of the phonon number by $E=\hbar\nu_{T}(\left\langle
a^{\dag}a\right\rangle +1/2)$. The expectation value for the number operator
can be read off from the second moments of the Wigner function:%
\begin{equation}
\left\langle a^{\dag}a\right\rangle \equiv\left\langle n\right\rangle
=\left\langle \bar{z}^{2}\right\rangle _{W}+\left\langle \bar{p}%
^{2}\right\rangle _{W}-\frac{1}{2}%
\end{equation}
We will calculate this quantity for different choices of parameters in the
following subsections.

\subsection{Cold damping}

\label{sec:colddamping}

In this subsection we show results for $\phi=-\pi/2$ and $\delta=0$, i.e. the
center of the band pass filter is set exactly to the trap frequency. As
derived in Appendix \ref{sec:fpe_solution} the number expectation value for
the steady state in this case is given by%
\begin{equation}
\left\langle n\right\rangle _{ss}=\frac{N+\frac{1}{2}\eta\tilde{\gamma}\left(
2N-1\right)  G+\frac{1}{8}\tilde{\gamma}G^{2}}{1+2\eta\tilde{\gamma}%
G}\label{eq:nss_colddamping}%
\end{equation}

\begin{figure}[ptb]
\centering
\includegraphics[width=.4\textwidth]{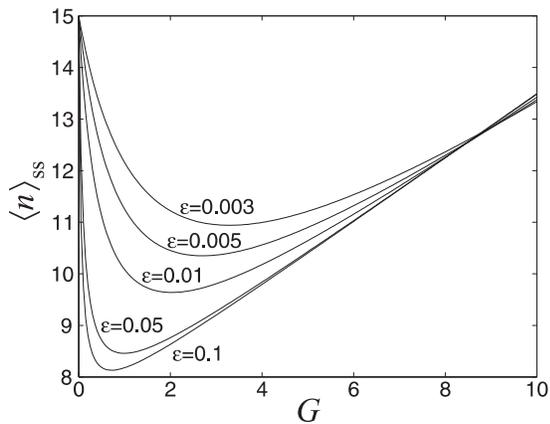}\caption{Number
expectation value for variable gain and solid angle fraction for $\eta=0.1 $,
$\phi=-\pi/2$, $\delta=0$ and $N=15$.}%
\label{fig:varepsilon}%
\end{figure}

Taking the gain $G=0$ yields $\left\langle n\right\rangle _{ss}=N$, i.e. if we
do not use the feedback current to influence the ion, the steady state
occupation will be the one for standard laser cooling. We see that the slope
of the occupation number is negative at $G=0$, i.e.
\begin{equation}
\partial\left\langle n\right\rangle _{ss}/\partial G|_{G=0}=-\tilde{\gamma
}\eta(2N+1)/2<0,
\end{equation}
and for $G\rightarrow\infty$ it diverges (note that in our model $G\gamma$ has to be smaller than $B$). Thus our theory yields a
non-vanishing optimal gain $G_{\mathrm{min }}$ for which the occupation number
has a minimum smaller than $N$,%
\begin{equation}
G_{\mathrm{min }}=\frac{\sqrt{1+8(2N+1)\eta^{2}\tilde{\gamma}}-1}{2\eta
\tilde{\gamma}}.
\end{equation}
Inserting this into Eq.~(\ref{eq:nss_colddamping}) yields an expression for
the minimal occupation number:%
\begin{equation}
\left\langle n\right\rangle _{\mathrm{min }}=\frac{4(2N-1)\eta^{2}%
\tilde{\gamma}-1+\sqrt{1+8(2N+1)\eta^{2}\tilde{\gamma}}}{16\eta^{2}%
\tilde{\gamma}}.\label{eq:cd_ssnumber}%
\end{equation}

\begin{figure}[ptb]
\includegraphics[width=.4\textwidth]{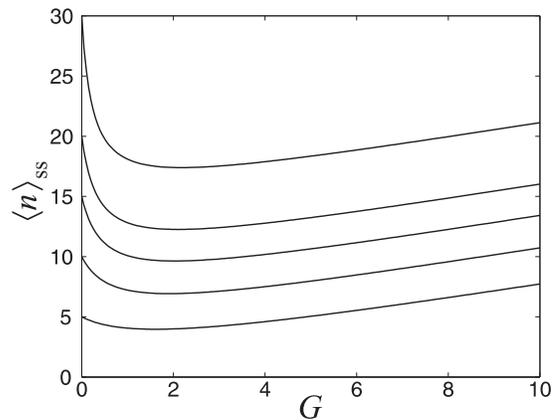}\caption{Number
expectation value for variable gain and steady state occupation number for
$\eta=0.1 $, $\phi=-\pi/2$, $\delta=0$ and $\varepsilon=0.01$. $N$ is given by
the values at $G=0$ of the different curves, from bottom to top, $N=5, 10, 15,
20, 30$.}%
\label{fig:varN}%
\end{figure}

With increasing solid angle $\varepsilon$ we collect more information about
the motional state of the system and hence the minimum $\langle n\rangle_{ss}$
is expected to decrease, which is shown in Fig.~\ref{fig:varepsilon}. With
increasing $\varepsilon$ the optimal gain is decreasing, because the feedback
noise term is growing with $G^{2}$ while the damping term is linear in $G$.

We show in Fig.~\ref{fig:varN} the decrease in the steady state phonon number with the gain. The relative decrease is larger with a higher laser cooling steady
state phonon number $N$. For lower $N$, the mirror decay rate $\gamma\propto N$ is smaller and thus we get less information about the motional state of the
atom, which limits the feedback cooling.

We will now expand $\left\langle n\right\rangle _{ss}$ in the limit of large
($N\gg1$) occupation numbers.\ For a series expansion of (\ref{eq:cd_ssnumber}%
) the formal expansion parameter is $N\sqrt{\varepsilon}$, thus an expansion
in the (usually also small) $\varepsilon$ is only possible for very low $N$.
We make an expansion for large $N$ in the opposite limit (Doppler limit),
while the condition $N\sqrt{\varepsilon}\gg1$ has to be satisfied. $N$ can be
tuned with e.g. with the laser detuning $\Delta_{L}$. Then the minimal
occupation number approximately reads%
\begin{equation}
\left\langle n\right\rangle _{\mathrm{min }}=\frac{N}{2}+4\sqrt{\frac
{1+\alpha}{\varepsilon}}-\frac{1+\alpha}{N\varepsilon}%
,\label{eq:nminphasepihalfNinf}%
\end{equation}
which implies that for a sufficiently large collection angle the minimal
obtainable phonon number is above $N/2$ and thus feedback cooling alone cannot
give a steady state. The reduction in the energy of the ion with time is due
to the reduction in $\left\langle \bar{p}^{2}\right\rangle _{W}$, while
$\left\langle \bar{z}^{2}\right\rangle _{W}$ is constant, as is shown in the
time evolution in Fig.~\ref{fig:variancesdelta0}. Thus the Wigner function for
the steady state will not be rotationally invariant, but \textquotedblleft
classically squeezed\textquotedblright\ in the momentum direction.

A phase space picture can demonstrate the action of the feedback on the system
state (see Fig.~\ref{fig:ps}(a)). By feeding back a linear force $f$ to the
ion, we effectively apply a unitary operator of the form
\begin{equation}
U(t)\sim\exp\left(  -\mathrm{i}fxt\right)  .\label{eq:f_unitary}%
\end{equation}
This operator acts as a momentum kick on a state with a magnitude proportional
to the momentum, which we have chosen by setting $\phi=-\pi/2$. The points in
the Wigner function will tend towards the x-axis, while the diffusion term
will counteract the feedback term, leading to a steady state Wigner function.

The difference in the position and momentum variance can be quantified; we
will give an expression for the amount of \textquotedblleft
squeezing\textquotedblright, i.e. the ratio between the two half-axis of the
error-ellipse for the Wigner function in phase space is obtained by rotating
the axes of the ellipse:
\begin{subequations}
\begin{align}
r_{\sigma} &  \equiv\frac{\text{semiminor axis}}{\text{semimajor axis}}%
=\frac{1-f}{1+f}\label{eq:squeezingparam}\\
f &  \equiv\frac{\sqrt{\left(  \sigma_{zz}-\sigma_{pp}\right)  ^{2}%
+4\sigma_{zp}^{2}}}{\sigma_{zz}+\sigma_{pp}}%
\end{align}

Here $\sigma_{zz}=\left\langle \bar{z}^{2}\right\rangle _{W}-\left\langle
\bar{z}\right\rangle _{W}^{2}$ and $\sigma_{pp}=\left\langle \bar{p}%
^{2}\right\rangle _{W}-\left\langle \bar{p}\right\rangle _{W}^{2}$ are the
variances of position and momentum, respectively, and $\sigma_{zp}%
=\left\langle \bar{z}\bar{p}\right\rangle _{W}-\left\langle \bar
{z}\right\rangle _{W}\left\langle \bar{p}\right\rangle _{W}$. As mentioned,
due to the affection of only the $\sigma_{pp}$ component, $\sigma_{zp}=0$ in
the case $\phi=-\pi/2$. The range of the squeezing parameter is $0<r_{\sigma
}\leq1$, where a small value corresponds to strong squeezing and for
$r_{\sigma}=1$ the state is symmetric.

The time dependent Fokker-Planck equation is solvable analytically and the
timescale of the cooling process is given by the eigenvalues of the drift
matrix (\ref{eq:kappamatrix}), which are in this case $\Gamma_{\mathrm{eff}} $
and $\Gamma_{\mathrm{eff}}+2\eta\gamma G$ corresponding to the usual Doppler
cooling and the feedback cooling. This shows that the feedback cooling happens
on a timescale faster than laser cooling alone.

\begin{figure}[ptb]
\centering
\includegraphics[width=.37\textwidth]{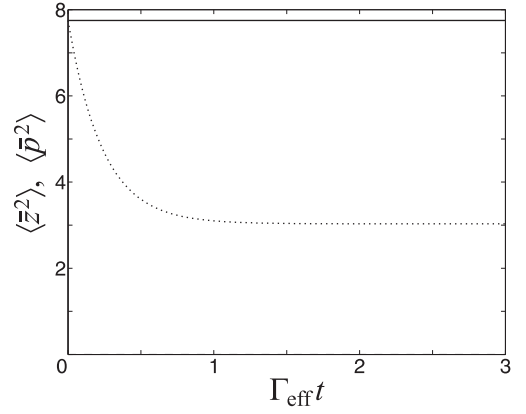}\caption{Time
evolution of the variances for $N=15$, $\tilde{\delta}=0$,
$\varepsilon =0.006$, $\eta=0.06$, $\phi=-\pi/2$ and
$G\approx1.51$ The solid line is the variance of the position
$\langle\bar{z}^{2}\rangle_{W}$ and the dotted line is the
variance of the momentum $\langle\bar{p}^{2}\rangle_{W}$.}
\label{fig:variancesdelta0}
\end{figure}

\subsection{Variable feedback phase}

For a phase $\phi\neq-\pi/2$, the magnitude of the feedback force is
proportional to the projection of the momentum on an other rotated axis in
phase space. We have pointed out in Eq.~(\ref{eq:f_unitary}) that the action
of the linear force (shifted trap) is always a momentum kick. Thus the
particle will always be \textquotedblleft kicked too hard\textquotedblright%
\ or not hard enough towards the phase space center. We will calculate the
regions of stability where the feedback can still lead to a steady state. Such
a steady state will only form if the both eigenvalues of the matrix $\kappa$
are positive. One eigenvalue of this matrix is $\Gamma_{\mathrm{eff}}$ for
arbitrary $\phi$, giving again the usual Doppler cooling, and the other
eigenvalue is $\Gamma_{\mathrm{eff}}-2G\eta\gamma\sin\phi$, which is always
positive for negative angles. For positive angles $\phi>0$, the gain has to
fulfill the condition $G<\Gamma_{\mathrm{eff}}/2\gamma\eta\sin\phi$. If this
condition is satisfied, a steady \ state number expectation value exists and
reads:
\end{subequations}
\begin{align}
\left\langle n\right\rangle _{ss} &  =\left[  \left(  1-\eta\tilde{\gamma
}G\sin\phi\right)  \left(  1-2\eta\tilde{\gamma}G\sin\phi\right)  \right]
^{-1}\times\label{eq:nssvarphase}\\
&  \times\lbrack N+\frac{1}{2}(4N-1)\eta\tilde{\gamma}G\sin\phi+\nonumber\\
&  +\frac{1}{8}\tilde{\gamma}G^{2}\left(  1+4\tilde{\gamma}\eta^{2}\left(
2N+1-2\sin^{2}\phi\right)  \right)  -\nonumber\\
&  -\eta\tilde{\gamma}^{2}G^{3}\sin\phi].\nonumber
\end{align}

\begin{figure}[ptb]
\centering
\includegraphics[width=.4\textwidth]{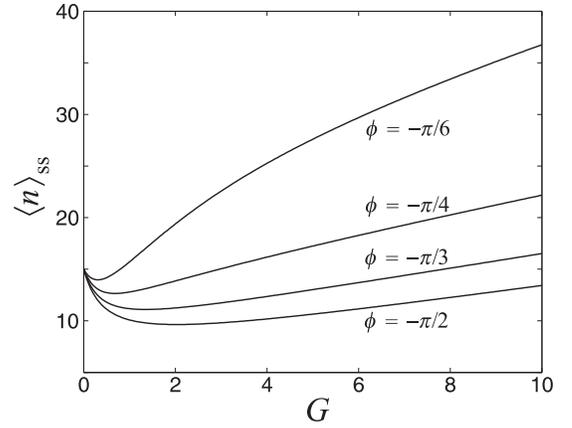}\caption{Number
expectation value for different feedback phases. Here $\delta=0$, thus the
expectation values for $\phi\rightarrow\pi-\phi$ yield the same results. Other
parameters: $N=15$, $\varepsilon=0.01$ and $\eta=0.1$.}%
\label{fig:varphi}%
\end{figure}

From Eq.~(\ref{eq:nssvarphase}) we can see that an energy decrease via
feedback cooling is only possible for angles $-\pi<\phi<0$ by calculating the
slope $\partial\left\langle n\right\rangle _{ss}/\partial G|_{G=0}$. Because
Eq.\thinspace(\ref{eq:nssvarphase}) is of higher order in $G$ than the
equation we had for $\phi=-\pi/2$, (\ref{eq:nss_colddamping}), we will not
give an analytical solution for the minimal gain and number occupation here.
We also find that for $\phi\neq-\pi/2$ the optimal occupation number is higher
than $\phi=-\pi/2$ (compare related studies in \cite{VitaliPhase}). The steady state occupation number for varying $\phi$ as a function of the gain
is plotted in Fig.~\ref{fig:varphi}, where we can see
that for non-optimal phases the range of $G$ for $\left\langle n\right\rangle
_{ss}<N$ is shrinking.

For the special case of $\phi=\pi$ or $\phi=0$, no cooling can be observed any
more and the number expectation value is quadratic in $G$. In principle, a
steady state with $\left\langle n\right\rangle _{ss}>N$ always exists with
\begin{equation}
\left\langle n\right\rangle _{ss}=N+\frac{1}{8}\tilde{\gamma}G^{2}\left(
1+4\eta^{2}\tilde{\gamma}\left(  2N+1\right)  \right)  .
\end{equation}
The more interesting feature of the $\phi=\pi$ case is that in the master
equation (\ref{eq:masterequation_final}) the feedback term (second term)
reduces in a rotating wave approximation to a Hamiltonian term of the form
$-\mathrm{i}\left[  \Delta\nu a^{\dag}a,\rho\right]  $. For this case we
observe a small shift $\Delta\nu$ in the frequency of the trap linearly
proportional to the gain. In this paper, we have not discussed the detailed
experimental setup used to apply the force to the ion, which would be
necessary for the knowledge of the exact forces acting on the ion. For
$\phi=\pi$ one can measure the frequency shift in the location of the sideband
and determine the conversion factor from the gain parameter $G$ used in this
paper and an experimental gain factor, which might be the real electronic gain
in the feedback loop.

\subsection{Rotation in Phase Space}

\begin{figure}[ptb]
\centering
\includegraphics[width=.37\textwidth]{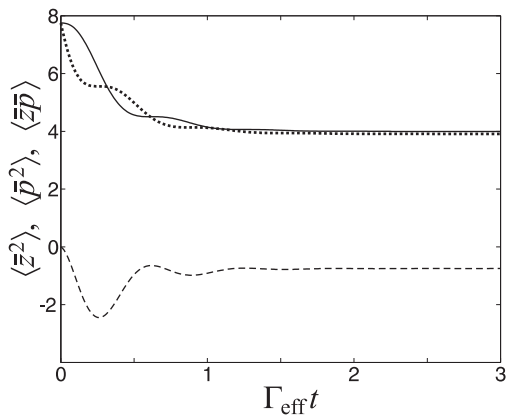}\caption{Time
evolution of variances $\langle\bar z^{2}\rangle_{W}$,
$\langle\bar p^{2}\rangle_{W}$ and $\langle\bar z\bar
p\rangle_{W}$ for $N=15$, $\tilde\delta=5$, $\varepsilon=0.006$,
$\eta=0.06$, $\phi\approx-85$ degrees and $G\approx2.03$ The solid
line is the variance of the position $\langle\bar
z^{2}\rangle_{W}$, the dotted line is the variance of the momentum
$\langle\bar p^{2}\rangle_{W}$ and the dashed line is $\langle\bar
z\bar
p\rangle_{W}$.}%
\label{fig:variancesdelta5}%
\end{figure}

\begin{figure}[ptb]
\centering
\includegraphics[width=.23\textwidth]{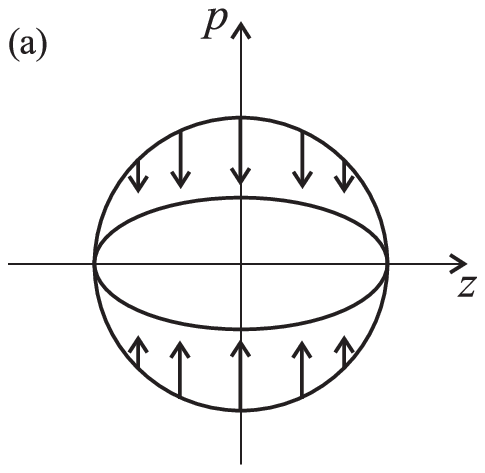}
\includegraphics[width=.23\textwidth]{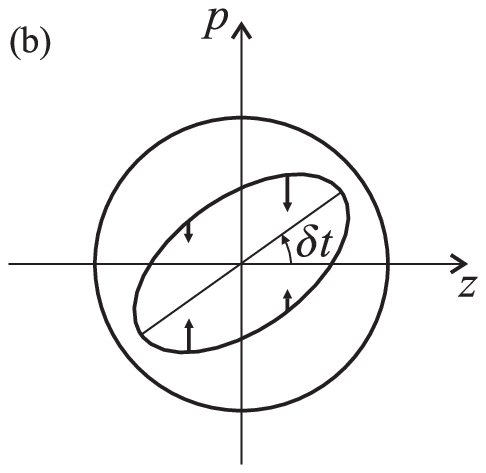}\caption{Schematic
drawing of the Wigner function error ellipses for the initial thermal state
(circle) and the feedback-cooled state (ellipse). In (a), $\phi=-\pi/2$ and
$\delta=0$, note that the position variance stays constant while the momentum
variance is decreased. The action of the force is always a kick in the
momentum direction and the force is proportional to the averaged momentum. In
(b), $\delta\neq0 $ and $\phi\approx-\pi/2$, here the Wigner function is
rotating and both variances are damped, resulting in lower energies.}%
\label{fig:ps}%
\end{figure}

We have shown that the phase $\phi=-\pi/2$ we chose leads to the lowest energy
of the motional state of the ion. The variance for the position operator
$\left\langle \bar{z}^{2}\right\rangle _{W}=\left(  2N+1\right)  /4$ remains
constant with time as shown e.g. in Fig.~\ref{fig:variancesdelta0}, thus
posing a lower limit to the obtainable energy. The detuning $\delta$\ of the
local oscillator in the feedback loop from the trap frequency creates a
tunable slow rotation of the (interaction picture) Wigner function in phase
space. This results in \textquotedblleft squeezing\textquotedblright\ of all
quadrature components (see Fig.~\ref{fig:ps}(b)), and the Wigner function can
regain a symmetric shape. Of course the timescale for this rotation has to be
much slower than the filter bandwidth $B$.

\begin{figure}[ptb]
\includegraphics[width=.4\textwidth]{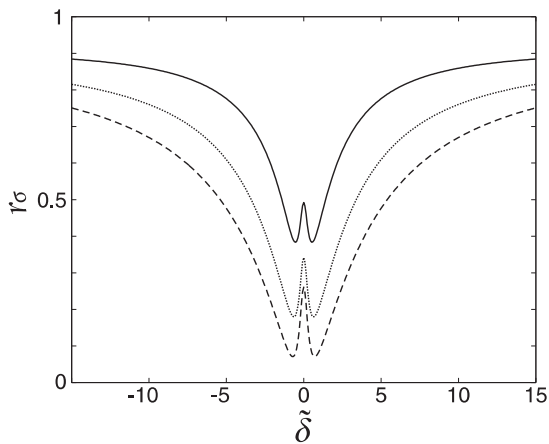}\caption{Squeezing
parameter $r_{{}}\sigma$ for the optimal phase at a given detuning and the
minimal energy state with $\varepsilon=0.006$, $\eta=0.06$ and with the
variable parameter $N=10$ (solid line), $N=17$ (dotted line) and $N=24$
(dashed line).}%
\label{fig:squeezeparams}%
\end{figure}

For the time evolution of the variances, the effect of the detuning is
illustrated in Fig.~\ref{fig:variancesdelta5}. We see the time evolution of an
initially thermal (symmetric) state with an occupation number of $N$. In
contrast to Fig.~\ref{fig:variancesdelta0} the width of the Wigner function in
the momentum and the position space are alternately decreased until they reach
the new feedback steady value. For a larger detunings the two variances
decrease equally in time and energetically lower states can be reached.
\begin{figure}[ptb]
\centering
\includegraphics[width=.4\textwidth]{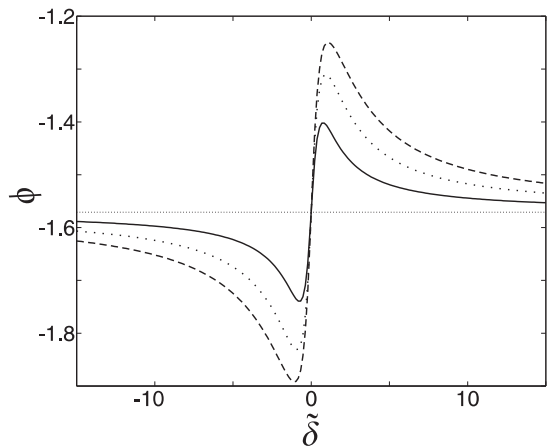}\caption{Optimal phase as
function of the detuning with $\eta=0.1$, $\varepsilon=0.006$ and $N=10$
(solid line), $N=17$ (dotted line) and $N=24$ (dashed line). The weak dotted
line marks $\phi=-\pi/2$.}%
\label{fig:minphi}%
\end{figure}For a rotation of the Wigner function with the frequency $\delta$,
we have to compare this rotation timescale with the cooling timescale $\gamma
$. For $\gamma$ comparable to $\delta$ the optimal phase is is shifted with
respect to $-\pi/2$ because the Wigner function is rotating in phase space
during the cooling time. When the detuning is much larger than the cooling
rate, the Wigner function ellipse direction will not be resolved during the
cooling time and thus the optimal phase returns to $-\pi/2$. By numerical
optimization (Fig.~\ref{fig:minphi}) we find that the optimal phase is shifted
from $-\pi/2$ asymmetrically with respect to the detuning $\delta$. It reaches
it's maximum excursion for a value of $\delta/\Gamma_{\mathrm{eff}}\approx1$,
for higher detunings the optimal phase approaches $-\pi/2$ again. For these
optimal values, we plot in Fig.~\ref{fig:squeezeparams} the squeezing
parameter $r_{\sigma}$ (\ref{eq:squeezingparam}), which is one for a symmetric
Gaussian state. We see that the state at no detuning is \textquotedblleft
classically squeezed\textquotedblright\ as we already mentioned in subsection
\ref{sec:colddamping} and the squeezing increases up to $\tilde{\delta}%
\approx1$, then upon approaching $\tilde{\delta}\rightarrow\infty$ the
squeezing parameter approaches one, and the state is thermal.

For increasing $\delta$, we also show that the number expectation value is
decreasing. We will not give an analytic expression for $\left\langle
n\right\rangle _{ss}$ for an arbitrary $\delta$ here. We merely calculate the
minimal number of phonons in the limit of large $\delta$. For this we require
an additional separation of the timescales between the effective feedback
cooling rate and the detuning, while the other timescale inequalities still
hold:%
\begin{equation}
\gamma\ll\delta\ll B.
\end{equation}
With these new conditions we take the detuning $\delta\rightarrow\infty$,
where the optimal feedback phase is again $\phi=-\pi/2$, and get for the
occupation number:%
\begin{equation}
\left\langle n\right\rangle _{ss}=\frac{N-\frac{1}{2}\eta\tilde{\gamma}%
G+\frac{1}{8}\tilde{\gamma}G^{2}}{1+\eta\tilde{\gamma}G}.
\end{equation}
The minimal occupation number for the same limit we took in deriving
Eq.~(\ref{eq:nminphasepihalfNinf}) we get for $N\gg1$:%
\begin{equation}
\left\langle n\right\rangle _{\mathrm{min }}\approx\sqrt{\frac{1+\alpha
}{2\varepsilon}}-\frac{1}{2}-\frac{2(1+\alpha)}{8\varepsilon N}.
\end{equation}
This expression does not include the large term $N/2$ any more and thus the
obtainable energy for large $N$ has an upper bound which is independent of $N
$, thus feedback cooling alone can give a thermal (symmetric) state with a
temperature below the Doppler temperature.

\section{Conclusion}

In this paper we have studied quantum feedback cooling of a trapped ion in
front of a mirror. This work is motivated by recent experiments
\cite{BushevThesis}, and -- as shown in \cite{FeedbackCoolingExp} -- provides
a quantitative understanding of the experimental results.

In the setup discussed in this paper the final temperatures are limited by the
collection efficiency, $\varepsilon$, and the constant scattering of photons
for the position measurement. This combination of heating due to the recoil,
and laser cooling due to the red detuning of the laser leads to a steady state
temperature (Doppler limit). The effect of quantum feedback cooling is studied
as an additional cooling mechanism on top of the ongoing laser cooling. For
the experimentally relevant parameters this leads to sub-Doppler cooling, but
it seems difficult to achieve ground state cooling in the trap along these
lines. As shown in a parallel publication \cite{Rabl}, we can devise a purely
\emph{dispersive} and thus non-invasive readout of the velocity of the trapped
ion based on the variation of the index of refraction with velocity, i.e.
based on electromagnetically induced transparency. Such a scheme allows, under
idealized conditions, ground state cooling of the ion purely by quantum feedback.

\begin{acknowledgments}
The authors thank R. Blatt, F. Dubin, J. Eschner, and D. Rotter
for discussions which motivated the present work. Research at the
University of Innsbruck is supported by the Austrian Science
Foundation, EU projects and the Institute of Quantum Information.
\end{acknowledgments}

\appendix

\section{Derivation of the Quantum Stochastic Master
Equation~(\ref{eq:qsme_general})}

\label{sec:meq_derivation} Starting from the stochastic master equation
(\ref{eq:ItoQSSE}) we define the reduced density matrix $\hat{W}%
(t)\equiv\operatorname{Tr}_{b}\left\{  \left\vert \Psi(t)\right\rangle
\left\langle \Psi(t)\right\vert \right\}  $. Note that $\hat{W}(t)$ is now a
trace-class operator for the internal electronic, the motional and the mirror
mode bath degrees of freedom. We calculate
\begin{equation}
d\hat{W}(t)=\operatorname{Tr}_{b}\left\{  \left\vert \Psi(t+dt)\right\rangle
\left\langle \Psi(t+dt)\right\vert -\left\vert \Psi(t)\right\rangle
\left\langle \Psi(t)\right\vert \right\}
\end{equation}
by inserting $\left\vert \Psi(t+dt)\right\rangle =\left\vert \Psi
(t)\right\rangle +d\left\vert \Psi(t)\right\rangle $ from
Eq.~(\ref{eq:ItoQSSE}). Using the Ito rules $dB_{u}(t)dB_{u^{\prime}}^{\dag
}(t)=\delta(u-u^{\prime})dt$, and cyclic property of the trace for background
bath operators, all terms of the form
\begin{align}
& \operatorname{Tr}_{b}\left\{  dB_{u}^{\dag}(t)\left\vert \Psi
(t)\right\rangle \left\langle \Psi(t)\right\vert \right\}  =\\
& =\operatorname{Tr}_{b}\left\{  \left\vert \Psi(t)\right\rangle \left\langle
\Psi(t)\right\vert dB_{u}(t)\right\}  =0\nonumber
\end{align}
vanish because the initial bath state is the vacuum state. With these rules we
obtain Eq.~(\ref{eq:qsme_general}).

\section{Adiabatic Elimination, Lamb-Dicke Limit and Laser Cooling}

\label{sec:AE_LDL_LC}

This appendix fills in the details of deriving the QSME
(\ref{eq:qsme_lambdicke}) from the QSSE (\ref{eq:ItoQSSE}) under the
assumption of weak driving and small Lamb-Dicke parameter. Note that we will
need to consider two different Lamb-Dicke parameters due to the exciting laser
which is not collinear with the $z$-axis. As in section \ref{sec:basics} we
denote $\tilde{\eta}\equiv\eta\sin\chi$. Inserting the ansatz
(\ref{eq:decompose}) into the QSSE (\ref{eq:ItoQSSE}) and transforming to an
interaction picture with respect to $H_{T}$ we get
\begin{align}
\left\vert \psi_{e}\left(  t\right)  \right\rangle = &  \frac{\mathrm{i}%
\Omega}{2}\left(  \frac{1-\frac{1}{2}\tilde{\eta}^{2}a^{\dag}a}{-\mathrm{i}%
\Delta_{L}+\frac{\Gamma}{2}}\right.  +\\
+ &  \left.  \frac{\mathrm{i}\tilde{\eta}a\mathrm{e}^{-\mathrm{i}\nu_{T}t}%
}{-\mathrm{i}\left(  \Delta_{L}-\nu_{T}\right)  +\frac{\Gamma}{2}}%
+\frac{\mathrm{i}\tilde{\eta}a^{\dag}\mathrm{e}^{\mathrm{i}\nu_{T}t}%
}{-\mathrm{i}\left(  \Delta_{L}+\nu_{T}\right)  +\frac{\Gamma}{2}}\right)
\left\vert \psi_{g}\left(  t\right)  \right\rangle .\nonumber
\end{align}
We insert this expression back into (\ref{eq:ItoQSSE}). We obtain%

\begin{align}
d\left\vert \psi_{g}\left(  t\right)  \right\rangle = &  \Biggl\{-\frac
{\Omega^{2}}{4}\Biggl[\frac{1}{-\mathrm{i}\Delta_{L}+\frac{\Gamma}{2}%
}+\mathcal{O}\left(  \mathrm{e}^{\pm2\mathrm{i}\nu_{T}t}\right)
+\label{eq:qsse_groundstate}\\
& + \frac{\tilde{\eta}^{2}a^{\dag}a}{\mathrm{i}\left(  \Delta_{L}-\nu
_{T}\right)  +\frac{\Gamma}{2}}+\frac{\tilde{\eta}^{2}aa^{\dag}}%
{\mathrm{i}\left(  \Delta_{L}+\nu_{T}\right)  +\frac{\Gamma}{2}}%
\Biggr]dt+\nonumber\\
+ &  \Biggl[\frac{\mathrm{i}\Omega}{2}\frac{1}{-\mathrm{i}\Delta_{L}%
+\frac{\Gamma}{2}}dC_{1}^{\dag}-\nonumber\\
- &  \left(  \frac{\mathrm{i}\tilde{\eta}a\mathrm{e}^{-\mathrm{i}\nu_{T}t}%
}{\mathrm{i}\left(  \Delta_{L}-\nu_{T}\right)  -\frac{\Gamma}{2}}%
+\frac{\mathrm{i}\tilde{\eta}a^{\dag}\mathrm{e}^{\mathrm{i}\nu_{T}t}%
}{\mathrm{i}\left(  \Delta_{L}+\nu_{T}\right)  -\frac{\Gamma}{2}}\right)
dC_{1}^{\dag}+\nonumber\\
+ &  \frac{\eta\left(  a\mathrm{e}^{-\mathrm{i}\nu_{T}t}+a^{\dag}%
\mathrm{e}^{\mathrm{i}\nu_{T}t}\right)  }{-\mathrm{i}\Delta_{L}+\frac{\Gamma
}{2}}dC_{2}^{\dag}\Biggr]\Biggr\}\left\vert \psi_{g}\left(  t\right)
\right\rangle \quad\mathrm{(I)}\nonumber
\end{align}
with
\begin{align}
dC_{1}^{\dag} &  \equiv\sqrt{\Gamma_{b}}\int du\,\sqrt{N\left(  u\right)
}dB_{u}^{\dag}+\sqrt{\Gamma_{m}}\sin(k_{eg}L)\,dB_{m}^{\dag},\\
dC_{2}^{\dag} &  \equiv-\mathrm{i}\sqrt{\Gamma_{b}}\int du\,\ \sqrt{N\left(
u\right)  }u\,dB_{u}^{\dag}+\cos(k_{eg}L)\,dB_{m}^{\dag}.
\end{align}
Consistent with the above approximations we neglect\ here and in the following
terms oscillating at twice the trap frequency $\nu_{T}$. Physically speaking,
the fourth line of Eq.\thinspace(\ref{eq:qsse_groundstate}) will correspond
together with third line to a heating and cooling term, and the last line
describes a diffusive term (cf. Fig. \ref{fig:coolheatdiffusion}).

Taking the trace over the background modes to define a reduced density
operator $\hat{w}(t)$ according to (\ref{eq:w}) we use the Ito rules, e.g.%
\[
\operatorname{Tr}_{b}\left\{  dB_{u}^{\dag}\left(  t\right)  \left\vert
\psi_{g}\left(  t\right)  \right\rangle \left\langle \psi_{g}\left(  t\right)
\right\vert dB_{u^{\prime}}\left(  t\right)  \right\}  =\delta\left(
u-u^{\prime}\right)  \rho\left(  t\right)  dt
\]
to derive Eq. (\ref{eq:qsme_lambdicke}).

\section{Homodyne photodetection and the diffusion approximation}

\label{sec:homodyne_diffusion}

As we have seen in Sec. \ref{sec:cond_meq}, the statistics of the detected
photons in the mirror mode are determined by the Poissonian stochastic
variable $dN_{c}(t)$. Like in homodyne detection, where a strong local
oscillator beats with the photodetection signal from a quantum system, an
elastic scattering term beats with the signal given by the coupling of the
light to the ion's motion (cf. Eq. (\ref{eq:meannumberdN})). The parameter
which gives the difference in the magnitudes of these terms is the Lamb-Dicke
parameter $\eta$. We split the stochastic variable $dN_{c}$ into a constant
(deterministic) part and a remaining stochastic part:%
\begin{equation}
dN_{c}(t)\equiv\frac{1}{2}\gamma dt+\eta dY_{c}\left(  t\right) .
\end{equation}
The stochastic expectation value of this equation is already known from Eq.
(\ref{eq:meannumberdN}):%
\begin{equation}
\left\langle dY_{c}(t)\right\rangle =\gamma\left\langle \tilde{z}\right\rangle
_{c}(t)dt.
\end{equation}
We check the distribution properties by calculating%
\begin{align}
dY_{c}^{2}\left(  t\right)   &  =\left(  \frac{dN_{c}\left(  t\right)
-\frac{1}{2}\gamma dt}{\eta}\right)  ^{2}=\frac{dN_{c}\left(  t\right)  }%
{\eta^{2}}=\\
&  =\frac{\frac{1}{2}\gamma dt+\eta dY_{c}\left(  t\right)  }{\eta^{2}%
}\overset{\eta\ll1}{\longrightarrow}\frac{\gamma}{2\eta^{2}}dt,\nonumber
\end{align}
which tells us that the stochastic variable has Gaussian properties, and thus
is associated with a white noise probability distribution. Thus $dY_{c}%
(t)=\sqrt{\gamma/2}/\eta\,dW(t)$ where $dW(t)$ is a Wiener increment.

The evolution of the system conditioned on measuring the photocurrent can be
seen by expanding the first bracket in the stochastic master equation
(\ref{eq:condmastereq}) to first order in the Lamb-Dicke parameter $\eta$, and
noting that the second bracket in (\ref{eq:condmastereq}) is
\begin{equation}
dN_{c}(t)-\left\langle dN_{c}(t)\right\rangle =\sqrt{\gamma/2}dW(t).
\end{equation}
Thus, using the formal derivative $\xi(t)=dW(t)/dt$ we obtain the conditioned
equation for the reduced density matrix, Eq. (\ref{eq:rhocond_homodyne_ito}).

\section{Equations of motion for the moments of the Wigner function}

\label{sec:fpe_solution}In Sec. \ref{sec:results} we use a Wigner function
representation for the density matrix and get an Fokker Planck equation
(\ref{eq:FPE}) equivalent to the master equation
Eq.~(\ref{eq:masterequation_final}) with the drift matrix
(\ref{eq:kappamatrix}) and the diffusion term (\ref{eq:diffusionterm}). The
equations of motion for the first and second moments of the Wigner function in
terms of the normalized position and momentum variables $\bar{z}=x_{1}$ and
$\bar{p}=x_{2}$, respectively, are:%

\begin{align}
\frac{\partial}{\partial t}\left\langle x_{i}(t)\right\rangle _{W}= &
-\sum_{j}\kappa_{ij}\left\langle x_{j}(t)\right\rangle _{W},\\
\frac{\partial}{\partial t}\left\langle x_{k}x_{l}\right\rangle _{W}= &
2D_{kl}+2D_{lk}-\\
&  -\sum_{j}\left[  \kappa_{kj}\left\langle x_{l}x_{j}\right\rangle
_{W}+\kappa_{lj}\left\langle x_{k}x_{j}\right\rangle _{W}\right]  .\nonumber
\end{align}
For a constant drift matrix, the equations for the first moments are trivial,
and if the eigenvalues of $\kappa$ are positive, the steady state value is
zero for both moments. We will therefore not concentrate on the first moments.
We will give the equations for the second moments which are relevant for the
number expectation value, and for this purpose we define a vector of second
moments
\begin{equation}
\mathbf{y}(t)=\left(  \left\langle \bar{z}^{2}(t)\right\rangle _{W}%
,\left\langle \bar{p}^{2}(t)\right\rangle _{W},\left\langle \bar{z}(t)\bar
{p}(t)\right\rangle _{W}\right)  ^{T}.
\end{equation}
We can write the equation of motion in a compact form as%
\begin{equation}
\mathbf{\dot{y}}(t)=M\mathbf{y}(t)+\mathbf{u}%
\end{equation}
where the evolution matrix is%
\begin{equation}
\frac{M}{\Gamma_{\mathrm{eff}}}=-\left(
\begin{array}
[c]{ccc}%
1 & 0 & -\tilde{\delta}\\
0 & 1-2\tilde{G}\sin\phi & \tilde{G}\cos\phi+\tilde{\delta}\\
2\tilde{G}\cos\phi+2\tilde{\delta} & -2\tilde{\delta} & 1-\tilde{G}\sin\phi
\end{array}
\right)  .
\end{equation}
Here $\tilde{G}\equiv G\eta\tilde{\gamma}$ and%
\begin{equation}
\mathbf{u}=\frac{\Gamma_{\mathrm{eff}}}{4}\left(  2N+1,2N+1+\frac
{\tilde{\gamma}}{2}G^{2},0\right)  ^{T}.
\end{equation}
The steady state results are obtained by setting $\mathbf{\dot{y}}(t)=0$,
which yields%
\begin{equation}
\mathbf{y}_{ss}=-M^{-1}\mathbf{u}%
\end{equation}
and we can calculate Eqs.~(\ref{eq:nss_colddamping}) and (\ref{eq:nssvarphase}%
) with $\left\langle n\right\rangle =y_{1}+y_{2}-1/2$.


\begin{thebibliography}{99}                                                                                               %
\bibitem {BlattRMP}D. Leibfried, R. Blatt, C. Monroe, and D. Wineland, Rev.
Mod. Phys \textbf{75}, 281 (2003), and references cited

\bibitem {WinelandCooling}
B. E. King, C. S. Wood, C. J. Myatt, Q. A. Turchette, D.
Leibfried, W. M. Itano, C. Monroe, and D. J. Wineland, Phys. Rev.
Lett. \textbf{81}, 1525 (1998)

\bibitem {EITCooling}C. F. Roos, D. Leibfried, A. Mundt, F. Schmidt-Kaler, J.
Eschner, R. Blatt, Phys. Rev. Lett. \textbf{85}, 5547 (2000) G.
Morigi, J. Eschner, and C. H. Keitel, Phys. Rev. Lett.
\textbf{85}, 4458 (2000)

\bibitem {CarmichaelBook}H. C. Carmichael, \textit{Statistical Methods in
Quantum Optics 1} (Springer, Berlin, 1999)

\bibitem {GZ}C. W. Gardiner and P. Zoller, \textit{Quantum Noise} (Springer,
Berlin, 2004), and references cited

\bibitem {WM_coll}H. M. Wiseman and G. J. Milburn, Phys. Rev. A \textbf{47},
642 (1993); H. M. Wiseman and G. J. Milburn, Phys. Rev. Lett. \textbf{70}, 548
(1993);  H. M. Wiseman, Phys. Rev. A \textbf{49}, 2133 (1994)

\bibitem {Feedback_Squeezing}H. M. Wiseman and G. J. Milburn, Phys. Rev. A
\textbf{49}, 1350 (1994)

\bibitem {FB_CQED}T. Fischer, P. Maunz, P. W. H. Pinkse, T. Puppe, and G.
Rempe, Phys. Rev. Lett. \textbf{88}, 163002 (2002); W. P. Smith, J. E. Reiner,
L. A. Orozco, S. Kuhr, and H. M. Wiseman, Phys. Rev. Lett. \textbf{89}, 133601
(2002); D. A. Steck, K. Jacobs, H. Mabuchi, T. Bhattacharya, and S. Habib,
Phys. Rev. Lett. \textbf{92}, 223004 (2004); P. Maunz, T. Puppe, I. Schuster,
N. Syassen, P. W. H. Pinkse, and G. Rempe, Nature \textbf{428}, 50 (2004) J.
McKeever, A. Boca, A. D. Boozer, R. Miller, J. R. Buck, A. Kuzmich, and H. J.
Kimble, Science \textbf{303}, 1992 (2004)

\bibitem {FB_Ion}J. A. Dunningham, H. M. Wiseman, and D. F. Walls, Phys. Rev.
A \textbf{55}, 1398 (1997); S. Mancini, D. Vitali, and P. Tombesi, Phys. Rev.
A \textbf{61}, 053404 (2000); J. Geremia, J. K. Stockton, and H. Mabuchi,
Science \textbf{304}, 270 (2004)

\bibitem {FB_Meso}S. Mancini, D. Vitali, and P. Tombesi, Phys. Rev. Lett.
\textbf{80}, 688 (1998); P. F. Cohadon, A. Heidmann, and M.
Pinard, Phys. Rev. Lett. \textbf{83}, 3174 (1999); A. Hopkins, K.
Jacobs, S. Habib, and K. Schwab, Phys. Rev. B \textbf{68}, 235328
(2003)

\bibitem {BushevThesis}P. Bushev, Ph.D. thesis, Universit\"at Innsbruck
(2004), url heart-c704.uibk.ac.at/dissertation/bushev\_diss.pdf

\bibitem {FeedbackCoolingExp}P. Bushev, D. Rotter, A. Wilson, F. Dubin, C.
Becher, J. Eschner, R. Blatt, V. Steixner, P. Rabl, and P. Zoller (2005), in preparation

\bibitem {MirrorColl}J. Eschner, Ch. Raab, F. Schmidt-Kaler, and R. Blatt,
Nature \textbf{413}, 495 (2001);  M. A. Wilson, P. Bushev, J. Eschner, F.
Schmidt-Kaler, C. Becher, R. Blatt, and U. Dorner, Phys. Rev. Lett.
\textbf{91}, 213602 (2003)

\bibitem {Rabl}P. Rabl, V. Steixner, and P.Zoller (2005), submitted for publication

\bibitem {CiracLaserCooling}J. I. Cirac, R. Blatt, P. Zoller, and W. D. Phillips, Phys. Rev. A
\textbf{46}, 2668 (1992), and references cited

\bibitem {Dorner02}U. Dorner and P. Zoller, Phys. Rev. A \textbf{66}, 023816 (2002)

\bibitem {EschnerNature}J. Eschner, C. Raab, F. Schmidt-Kaler, and R. Blatt,
Nature \textbf{413}, 495 (2001)

\bibitem {ResFluorescence}J. I. Cirac, R. Blatt, A. S. Parkins, and P. Zoller,
Phys. Rev. A \textbf{48}, 2169 (1993)

\bibitem {Risken}H. Risken, \textit{The Fokker-Planck equation} (Springer,
Berlin, 1989)

\bibitem{VitaliPhase}
S. Mancini, D. Vitali, and P. Tombesi, Phys. Rev. A \textbf{61},
053404 (2000)

\end{thebibliography}
\end{document}